# Simultaneous imaging of bidirectional guided waves enables synchronous probing of mechanical anisotropy, local blood pressure, and stress in arteries

Yuxuan Jiang [1], Guo-Yang Li [2,*], Keshuai Hu[1], Shiyu Ma [1], Yang Zheng [1], Mingwei Jiang [1], Zhaoyi Zhang [1], Xinyu Wang [3,4,5,*], Yanping Cao [1,*]

[1] Institute of Biomechanics and Medical Engineering, AML, Department of Engineering Mechanics, Tsinghua University, Beijing 100084, China.

[2] Department of Mechanics and Engineering Science, College of Engineering, Peking University, Beijing 100871, China.

[3] Department of Cardiology, Peking University Third Hospital, Beijing 100191, China.

[4] NHC Key Laboratory of Cardiovascular Molecular Biology and Regulatory Peptides, Peking University, Beijing 100191, China.

[5] Beijing Key Laboratory of Cardiovascular Receptors Research, Beijing 100191, China.

*Corresponding authors: lgy@pku.edu.cn (G.Y.L.); wangxinyu@bjmu.edu.cn (X.W); caoyanping@tsinghua.edu.cn (Y.C.)

**Abstract**

Arterial biomechanical indicators have long been recognized as fundamental contributors to the physiology and pathology of cardiovascular systems. Probing the multiple biomechanical parameters of arteries simultaneously at different time points within one cardiac cycle is of great importance but remains challenging. Here we report an ultrasound elastography method to quantify arterial anisotropic stiffness, mechanical stresses in arterial wall, and local blood pressure in a single measurement. With programmed acoustic radiation force, arterial axial and circumferential guided elastic waves were induced simultaneously and recorded at multiple time points within one cardiac cycle. Then a mechanical model incorporating acoustoelasticity and viscoelasticity of arteries was proposed to quantitatively predict the correlation of arterial guided elastic waves with arterial biomechanical parameters. Our experimental design and biomechanical model lead to an elastography method to interrogate the variation of blood pressure, arterial bidirectional stiffnesses and mechanical stresses in arterial walls with time. *In vivo* experiments were performed on healthy young, normotensive older and hypertensive older volunteers. The results demonstrate that the reported method can find applications in understanding aging of cardiovascular system and diagnosis of cardiovascular diseases.

**One-sentence teaser**: Ultrasound elastography simultaneously measures arterial anisotropic stiffness, local blood pressure, and bidirectional mechanical stress, offering insights into cardiovascular health and diseases.

## 1. Introduction

Arterial biomechanical indicators including arterial stiffness, blood pressure and mechanical stresses in arterial wall have long been recognized as fundamental contributors to the physiology and pathology of cardiovascular system. Arterial stiffness is profoundly affected by aging and various cardiovascular diseases (CVDs), including diabetes, hypertension and hypercholesterolemia (*1, 2*). Adversely, large-artery stiffening will cause isolated systolic hypertension, left ventricular dysfunction and target organ damage (*3*). Therefore, arterial stiffness has been used as a risk factor for various CVDs including hypertension, diabetes and atherosclerosis (*4-7*). Blood pressure is another key biomarker in cardiovascular system, its monitoring and management are essential for patients with hypertension (*8*). It has been recognized that the blood pressure and arterial stiffness have a complex interplay, leading to an insidious feedback loop in the progression of CVDs (*3*). Mechanical stresses in arterial wall stem from residual stresses required to guarantee the physiology functions of artery and prestresses caused by blood pressure, which are sensed by vascular cells and regulates the homeostasis of arteries (*9, 10*). Changes of mechanical stress will alter the homeostasis, leading to arterial remolding and disease progression (*11*). Excessive arterial stress can predispose to aortic dissection (*12*) and rupture of aneurysms (*13*). Given the importance of these biomechanical indicators for diagnosis and therapy of CVDs, their measurements *in vivo* have long been pursued over past years.

Arteries are fiber-reinforced materials and exhibit mechanical anisotropy (i.e. distinct variations in axial and circumferential stiffness) (*14*). Diverse pathologies have been linked to the changes of arterial anisotropy (*15*). In clinics, pulse wave velocity (*5*), arterial compliance and stiffness index (*16, 17*) enable the assessment of the equivalent stiffness of arteries in a specific direction. Characterizing the anisotropic stiffness of arteries *in vivo* remains challenging. Arterial guided wave elastography method developed recently holds promise to characterize arterial anisotropic stiffness (*18*). In this method, the arterial wall is stimulated by focused acoustic radiation forces and the induced elastic wave propagates in a guided way confined by the arterial wall, which can be detected with ultrafast ultrasound imaging. Imaging guided waves along the axial direction (*19-27*) or in cross-sectional plane (*28, 29*), arterial guided wave elastography possesses the capability to gauge arterial axial or circumferential stiffness, respectively. The methods reported in the literature image axial and circumferential guided waves separately by rotating the ultrasound probe (*30, 31*); therefore, it remains challenging to accurately measure the anisotropy in arterial stiffness because the arterial stiffness is spatial dependent and varies in cardiac cycles due to pulsation of blood pressure. To overcome this challenge, it is necessary to image the axial and circumferential guided waves simultaneously at the same position of artery.

Blood pressure leads to circumferential mechanical stress in arterial wall (*32, 33*) and axial stress is mainly attributed to the prestresses caused by the interaction of artery with surrounding soft tissues (*34, 35*). Emerging evidence indicates that axial stress significantly contributes to vascular homeostasis and aids in minimizing overall tissue stress in normal arteries (*11*). Existing methods for measuring arterial circumferential stress need to measure local blood pressure and corresponding arterial deformation, whereas the assessment of axial stresses requires the measurement of arterial anisotropic mechanical parameters and axial deformation (*34, 36*), which are difficult to achieve. Our previous work has shown that programmed elastic waves enable probing mechanical stress in soft materials without prior knowledge of constitutive models (*37*) and variation of local blood pressure is related to variation of guided waves (*20*). Therefore, local blood pressure and mechanical stresses in artery at different moments of a cardiac cycle are possible to be inferred provided the axial and circumferential guided waves at the same position of artery can be measured simultaneously.

In this study, we report a guided wave elastography method for arteries, in which the axial and circumferential guided waves are successfully generated with an exclusively programmed acoustic radiation force. We then develop an image processing algorithm and an acoustoelastic model incorporating tissue viscoelasticity to infer arterial anisotropic stiffness, local blood pressure and mechanical stresses in arterial wall in a single measurement. We conducted *in vivo* experiments on young ($n$ = 30), older ($n$ = 14) and hypertension volunteers ($n$ = 8) to demonstrate the potential use of our method in clinics. The capability to measure arterial anisotropic stiffness, local blood pressure and mechanical stresses in arterial wall simultaneously at a local position of artery at different time points of a cardiac cycle not only helps understand the relationship between these crucial biomechanical indicators, but facilitates their application in clinics for the diagnosis and therapy of cardiovascular diseases.

## 2. Results

### 2.1 Generating and imaging axial and circumferential guided waves simultaneously in the longitudinal ultrasound imaging view

Our approach was built on remote excitation of bidirectional guided waves with programmed acoustic radiation force (ARF) and imaging the wave motions in axial view of the human right common carotid arteries (CCA, see Fig. 1A). We utilized a linear ultrasonic transducer to apply ten transient focused ultrasound beams to the proximal side of the artery, sequentially from the anterior wall to the posterior wall with a supershear speed in a time span of ~0.5 ms, which results in a cone shape wave front (Fig. 1B). After a rest duration of ~0.3 ms, the same transducer started to perform plane wave imaging at a pulse repetition frequency of 10 kHz in a period of 4 ms (40 frames in total), ensuring that the imaging window covered both the anterior and posterior walls. The data were recorded in I/Q data and processed offline to obtain the tissue particle velocity fields, enabling the analysis of guided wave propagation. The aforementioned measurement was repeated 36 times, with a resting duration of ~65 ms between adjacent two measurements. Totally the measurements took ~2.5 s, spanning around 3 cardiac cycles (Fig. 1C). The electrocardiogram (ECG) signal was recorded synchronously in the measurements for reference.

Figure 1D shows the snap shots of the particle velocity map for a healthy young volunteer (27 years old, male). Figure 1E depicts the spatiotemporal map of the particle velocity extracted along the centerline of the anterior wall. The front of the axial guided wave is tracked by the red dash line (the slope denotes the axial wave velocity $c_a$). Interestingly, circumferential guided wave manifests itself in two minima in the time profile of the particle velocity at $z = 0$. We denote the corresponding time points as $t_1$ and $t_2$, respectively (Fig. 1E). The emergence of the circumferential guided wave in the axial view can be understood as follows: the circumferential guided wave that originates from the posterior wall follows a semi-circular path, arriving at the anterior wall at time $t_1$. Subsequently, at time $t_2$, the circumferential guided wave emanating from the anterior wall completes its journey, returning to its initial point. To verify this assumption, we performed a comparative experiment where only the anterior wall of the vessel was excited. In this case, no negative peak was observed at time $t_1$, and only a weak peak (due to strong attenuation in arteries) appeared at time $t_2$ (see Supplementary Note 1 and Fig. S1).

We performed finite element analysis (FEA) and phantom experiments to understand the effects of arterial biomechanics on the observed elastic wave propagation. The simulations turned out that the viscoelasticity makes arteries a low-pass filter that filers out high frequency elastic waves, resulting in a weak dispersion that enables accurate identification of the time $t_1$ by tracking the axial wave peaks (see Fig. 1F, Supplementary Note 2 and Fig. S2). Since the time $t_1$ corresponds to the arrival of the circumferential guided wave originating from the posterior wall, we can determine the circumferential wave speed $c_c$ through $c_c = \pi r_c/(t_1+t_c)$, where $t_c$ denotes the compensation time (0.8 ms, accounting for the duration of ARFs and Null stages, see Fig. 1D) and $r_c$ is the center radius of the artery. Differently, the axial wave speed $c_a$ is determined by tracking the peaks in the spatiotemporal map (see the dashed lines in Figs. 1E and F).

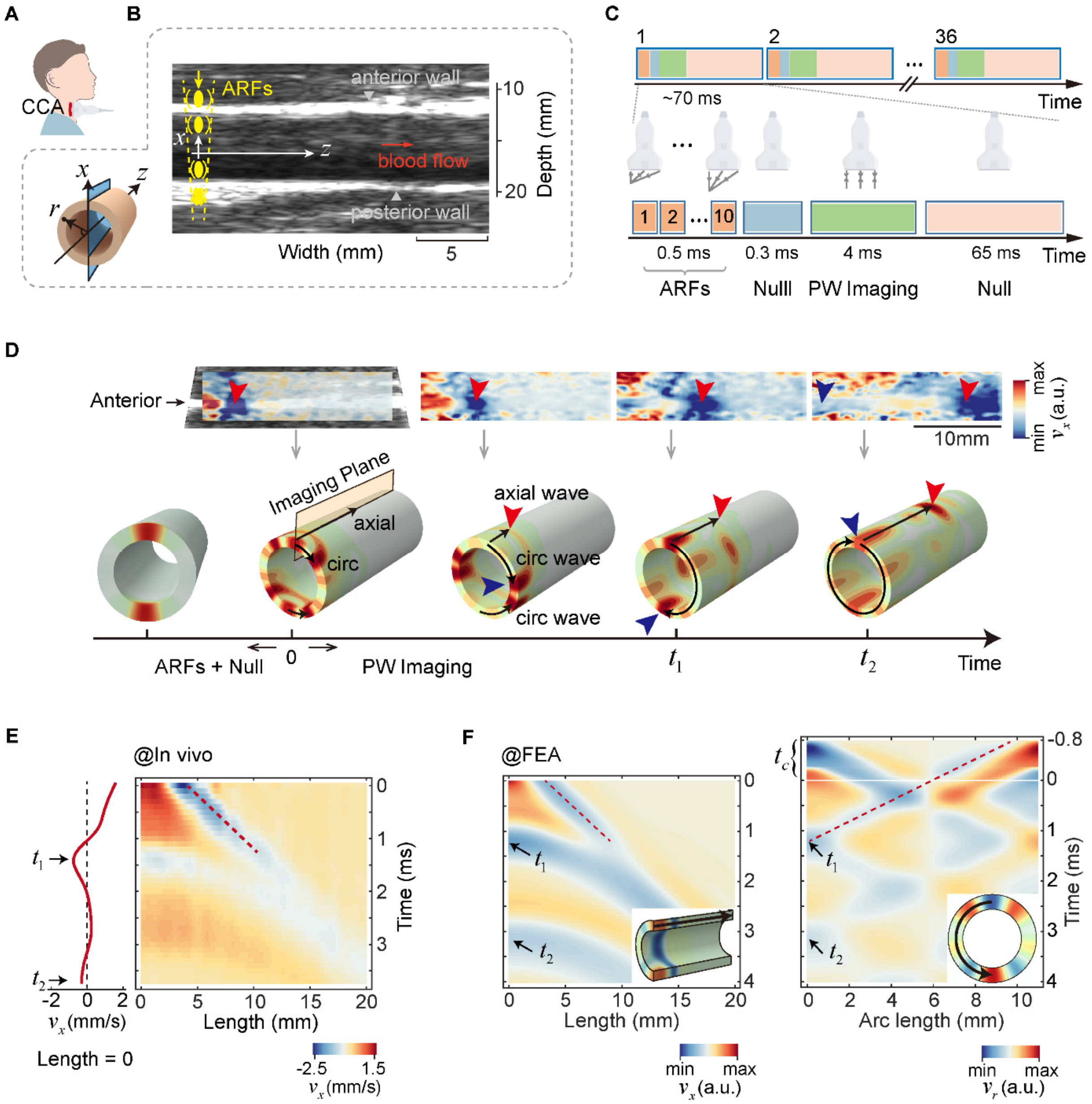

**Fig. 1. Excitation and measurement of axial and circumferential guided waves. (A)** Experiment set-up to measure human right CCAs. **(B)** Ultrasound imaging and excitation method. **(C)** Imaging sequence. PW: Plane Wave. **(D)** Particle velocity maps in the longitudinal view with the variation of time from the *in vivo* experiment and corresponding schematic to show the wave propagation. Circ: Circumferential. **(E)** Particle spatiotemporal velocity map extracted along the axial path on the anterior wall at end-diastole from the *in vivo* experiment. **(F)** Simulation of the bidirectional guided waves, including velocity map extracted along the axial path (left), and velocity map along the circumferential path (right).

## 2.2 Circumferential guided wave velocity as a robust index for probing continuous blood pressure

Figure 2A shows the dynamic variation of axial and circumferential wave velocities during cardiac cycles from a young healthy volunteer (27 years old, male). For reference, the figure also depicts the variations of arterial diameter and blood pressure, as well as the electrocardiogram (ECG). As expected, the axial and circumferential wave velocities vary simultaneously with diameters as well as the blood pressure. The right panel of Fig. 2A shows six snapshots of the particle velocity at specific cardiac phases. At the middle time of the systole (state ii) and around the dicrotic notch (state iv), the large motions of the arterial wall induced by pulse wave are obvious in the maps. Interestingly, we observed that the time phase $t_1$ varied inversely with the diameter; it reached the minimum when the diameter was at its maximum, and vice versa (Supplementary Fig. S3). Since the wave propagates a longer arc length in an artery with a greater radius, this observation suggests the change in wave velocity is more significant than the change in arterial geometry. Collectively, it is concluded that the increase in blood pressure enlarges both the axial and circumferential stiffness and stress of the artery, and in turn dramatically accelerates the wave propagation.

To test this conclusion, we recruited healthy young volunteers ($n$ = 30, 20 ± 2 years old), normotensive older volunteers ($n$ = 14, 51 ± 8 years old), and hypertensive older volunteers ($n$ = 8, 55 ± 5 years old) for the *in vivo* experiment. Figure 2B shows three representative particle velocity maps at end-diastole from young, older and hypertensive groups, as well as the corresponding bidirectional wave velocities in cardiac cycles. We observed that for most older volunteers (11 out of 14) and hypertensive volunteers (8 out of 8), the time phase $t_1$ was less pronounced on the anterior wall, whereas it could be clearly detected on the posterior wall (see details in Supplementary Fig. S4). This can be attributed to stronger ultrasound scattering and attenuation in older and hypertension volunteers, resulting in a much weaker acoustic radiation force applied on the posterior wall than that on the anterior wall (the ratio of acoustic radiation force on the posterior wall to the anterior wall could be less than 0.25 in the hypertensive group; see details in Supplementary Note 3). Consequently, a stronger circumferential signal is detected on the posterior wall, corresponding to guided waves originating from the anterior wall. Our simulations also verify that a dramatically greater force on the anterior wall than the posterior wall is needed to reproduce the experiments (Fig. 2C).

Figure 2D displays the statistical results of bidirectional wave velocities at diastole (corresponding to the minimum velocity within a single cardiac cycle) and systole (corresponding to the maximum velocity within a single cardiac cycle) in the young, older, and hypertensive groups, respectively (the values of wave velocities are listed in Table 1). To validate, we find the axial wave velocities agree well with those reported in previous studies (*20, 21*). In most cases, the axial and circumferential wave velocities exhibit no significant difference; except in the young group at diastole and the hypertensive group at systole ($p < 0.05$). The difference in axial and circumferential wave velocities is primarily attributed to multiple factors including the anisotropic arterial stiffness, mechanical stress and geometrical curvature. This point will be discussed in details in the following section with the help of a quantitative mechanical model.

Figure 2E compares the statistical results of bidirectional wave velocities at diastole, across the three groups. Both axial and circumferential wave velocities increase with aging, as both axial and circumferential waves in older volunteers (normotensive old and hypertensive old volunteers) travel faster than those in healthy young volunteers. Surprisingly, we find hypertension further elevates circumferential wave velocities, but not significantly changes the axial wave velocities. This may ascribe to the remodeling of arteries in response to the elevated circumferential stresses introduced by hypertensions, leading to the increases in circumferential arterial stiffnesses that accelerate the circumferential wave velocities. Figure 2F shows the data obtained at systole. Differently, both the aging and hypertension manifest themselves in the elevated wave velocities in the two directions. The greater axial velocities in hypertensive old volunteers than those of the normotensive old volunteers ($p < 0.01$) indicates that the systolic pressure not only affects the circumferential stiffness and stress but also increases those in the axial directions. These observations evidence that both axial and circumferential wave velocities are closely related to blood pressure and thus can be indicators to interrogate blood pressure locally.

As illustrated in Fig. 3A, the squares of both the axial and circumferential wave velocities from three volunteers are linearly correlated to the blood pressures. To validate, we also performed FEA, as shown in Fig. 3B. Importantly, this linear relation enables the quantitative measurement of blood pressure by using either axial or circumferential wave velocities. Despite the axial wave velocities have been reported to be useful in probing blood pressure (*20*), we find it is beneficial to employ circumferential wave velocities in terms of stability and robustness. We conducted a comparative experiment on a young volunteer (27 years old, male) involving two different postures. The volunteer was first instructed to sit upright with a forward gaze (referred to as the normal posture), and then to extend his neck fully by looking upward to the utmost (referred to as the craning posture), during which the ultrasound elastography was performed. The measurements were repeated three times for each posture. The left of Fig. 3C shows the particle velocity map at diastole in normal posture and the corresponding bidirectional wave velocities. The right of Fig. 3C shows the particle velocity map in craning posture and the corresponding bidirectional wave velocities. As shown, the circumferential wave velocity remains almost unchanged (relative change < 1%) while the axial wave velocity increases obviously when craning (relative increase ~17%). This can be attributed to the fact that craning leads to a more pronounced axial stretch of the CCA, while blood pressure remains stable during two successive and short-term experiments conducted under different postures. FEA in Fig. 3B also confirms this variation: when the axial stretch increases, axial wave velocities significantly increase, while circumferential wave velocities remain stable. As a result, circumferential wave velocities are less affected by changes in neck posture, making them a more robust and preferable means of measuring blood pressure.

Figure 3D shows a representative result of blood pressure measurement using the circumferential wave velocities, alongside applanation tonometry (*38*), which is considered the gold standard for noninvasive continuous blood pressure measurement. We adopted the Bland-Altman plot to assess the consistency between the two methods. The biases are −0.06 mmHg in average, with 95% limits of agreement (LoA) of −6.0 to 5.9 mmHg. The statistical results show that the consistency of the methods is good (B-A plots for other groups are shown in Supplementary Fig. S5). Figure 3E shows more results of blood pressure measurement among the three groups. In general, both waveform and amplitude of the blood pressure can be effectively captured by the elastography method. Compared to tonometry, this method eliminates the need for vascular compression, making it well-suited for long-term blood pressure monitoring. Additionally, unlike methods that infer blood pressure by measuring arterial geometry (*39, 40*), this method is less susceptible to posture variations and body movements, offering the potential for more accurate blood pressure measurements. Therefore, it holds promise as a clinical tool for noninvasive, continuous, and long-term blood pressure assessment in hypertensive patients.

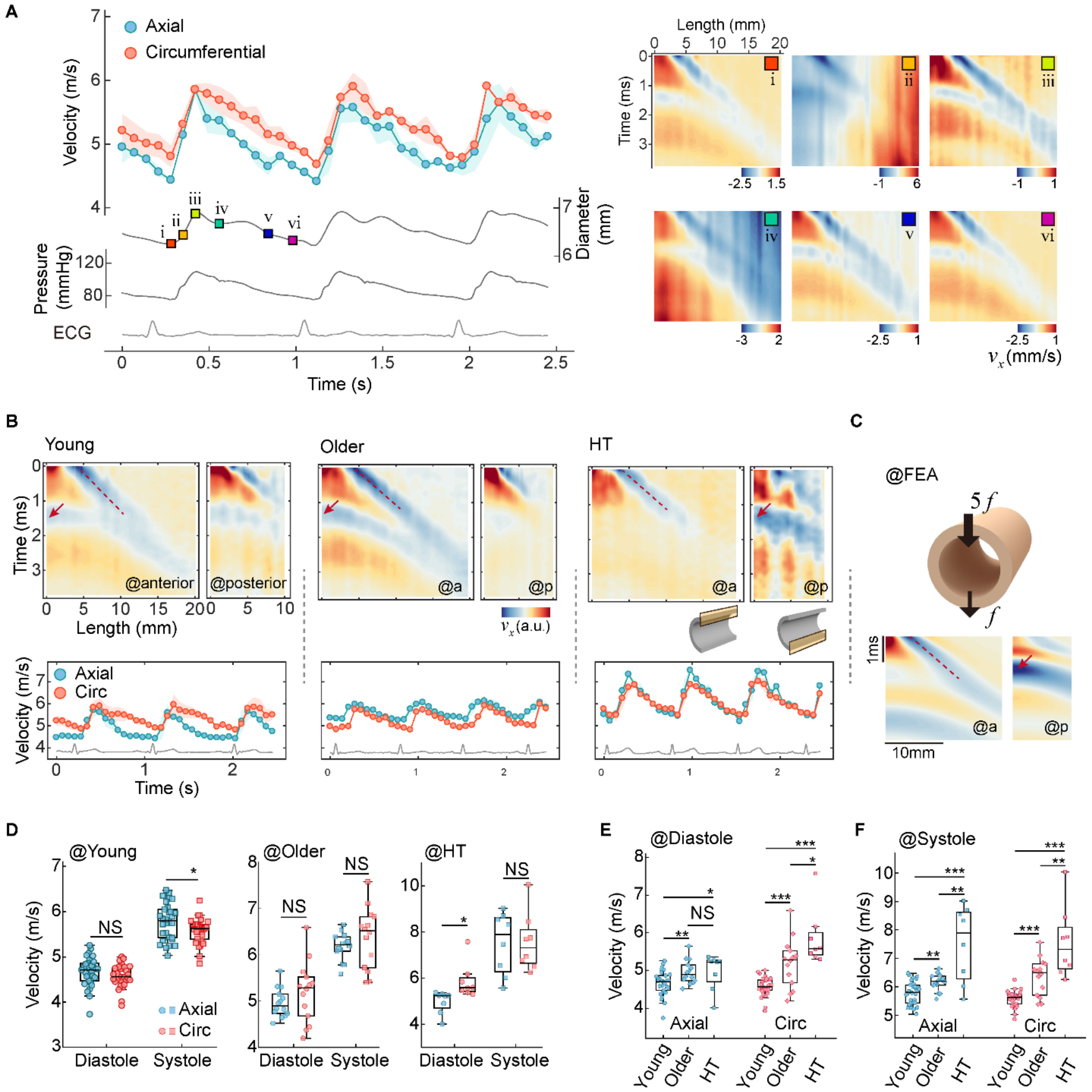

**Fig. 2. Dynamic variation of the bidirectional wave velocities in the young, older and hypertensive groups. (A)** Variation of the bidirectional wave velocities and spatiotemporal velocity maps in cardiac cycles from a young volunteer. The synchronous arterial diameter, blood pressure (measured using applanation tonometry), and ECG are also plotted. **(B)** Representative particle velocity maps extracted from the anterior and posterior walls at end-diastole for volunteers in the three groups, and their corresponding bidirectional wave velocities. **(C)** FEA to verify the difference of spatiotemporal maps extracted from the anterior and posterior walls. The acoustic radiation force (*f*) applied on the anterior wall is 5 times larger than that on the posterior wall. **(D)** Statistics of the bidirectional wave velocities at diastole and systole in the young, older and hypertensive groups, respectively. **(E)** Comparison of bidirectional wave velocities at diastole and **(F)** systole among the three groups.

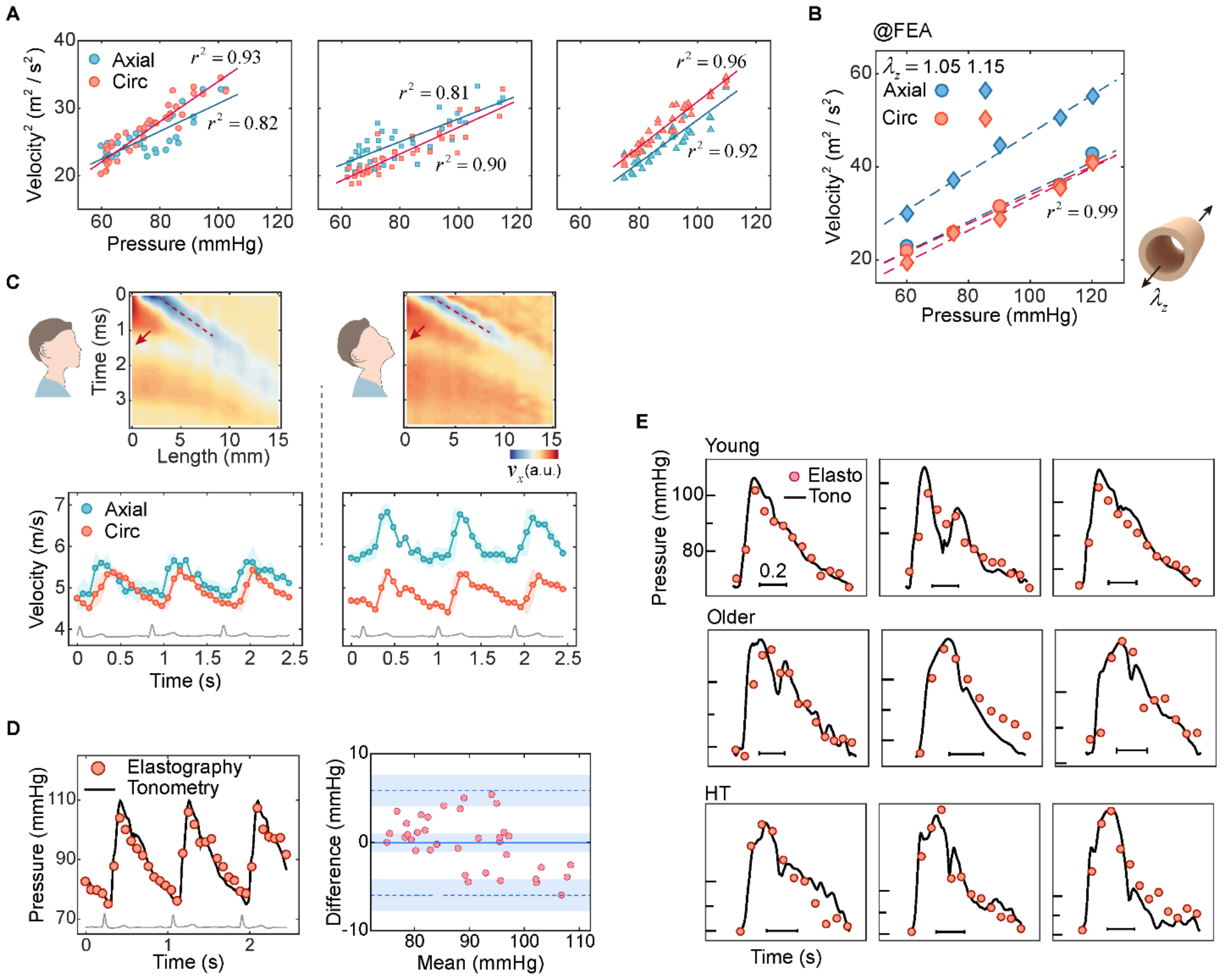

**Fig. 3**. **Blood pressure measurement. (A)** Relationship of the squares of the bidirectional wave velocities and blood pressure from three young volunteers. **(B)** FEA results to verify the linear relationship between the squares of bidirectional velocities and blood pressure, and the variation of bidirectional wave velocities in the two postures. **(C)** A comparative experiment conducted in two postures: normal posture (a forward gaze) and craning posture (stretching his/her neck). Top left: Particle velocity maps at end-diastole for the normal posture; top right: for the craning posture. Bottom left: Bidirectional wave velocities for the normal posture; bottom right: for the craning posture. **(D)** Blood pressure measurement of a young volunteer. The elastography method using circumferential wave velocities is compared to the applanation tonometry method; and B-A plot of the two methods. **(E)** Blood pressure measurement results from nine volunteers across the young, older, and hypertensive groups.

**Table 1. Results of bidirectional wave velocities in the three groups**

| | Young ($n$=30) | Older ($n$=14) | HT ($n$=8) |
|---|---|---|---|
| $c_{a,\mathrm{d}}$ (m/s) | 4.67±0.32 | 4.96±0.31 | 4.97±0.48 |
| $c_{a,\mathrm{s}}$ (m/s) | 5.78±0.40 | 6.15±0.30 | 7.52±1.33 |
| $c_{c,\mathrm{d}}$ (m/s) | 4.58±0.25 | 5.20±0.65 | 5.86±0.75 |
| $c_{c,\mathrm{s}}$ (m/s) | 5.58±0.28 | 6.35±0.66 | 7.55±1.25 |

## 2.3 Bidirectional guided wave imaging enables synchronous measurement of bidirectional stiffness and mechanical stress in arterial walls

It is well recognized that arterial biomechanics is highly nonlinear, resulting in a stress-dependent arterial stiffness. The experimental data suggest both arterial stress and stiffness play an important role in shear wave propagation. We therefore developed a mechanical model for bidirectional guided wave elastography. We noted that the duration of wave propagation (~ microsecond) is much shorter than the cardiac periodicity (~ second), and the deformation involved in wave propagation (~ micrometer) is much smaller than the deformation introduced by pulse wave motion (~ millimeter). It is reasonable to model the shear wave motion as an infinitesimal deformation imposed on a static finite deformation, which can be described by the following motion equation (*41*)

$$\nabla \cdot \mathbf{\Sigma} = \rho \mathbf{u}_{,tt} \tag{1}$$

where $\mathbf{\Sigma}$ denotes incremental stress induced by elastic waves. $\mathbf{u}$ denotes the displacement induced by elastic wave motions. $\rho$ is the density of arteries and $t$ is the time. The subscript comma indicates partial derivative of the variable. For harmonic waves, $\mathbf{u} = \mathbf{u}_0 e^{i(\mathbf{k}\cdot\mathbf{x} - \omega t)}$, where $\mathbf{u}_0$, $\mathbf{k}$ and $\omega$ denote wave amplitude, wave vector and angular frequency, respectively. For guided waves in arteries, we seek the relation between $\omega$ and wavenumber $k = |\mathbf{k}|$, which determines the dispersion relations and phase velocities. We model the arterial wall as a nonlinear viscoelastic hollow cylinder surrounded by fluid. Blood fills the interior, while the perivascular tissues are so soft that they can be approximated as fluid, as supported by our previous work (*42*). With these boundary conditions, we obtain the phase velocity of the axial guided waves $c_a^p$ as

$$c_a^p = \frac{\omega}{\mathrm{Re}(k)} \left[ 1 - \left( \frac{N}{r_c \, \mathrm{Re}(k)} \right)^2 \right]^{-\frac{1}{2}} \tag{2}$$

The relation between the angular frequency $\omega$ and the wavenumber $k$ implicitly depends on viscoelastic parameters of the artery $\alpha_a$, $\gamma$, $g$, $\tau$ (where $\alpha_a$ and $\gamma$ describe the arterial nonlinear elasticity, $g$ and $\tau$ describe the arterial viscosity) and wall thickness $h$ (see Eq. (9) in Methods, section 4.14). The additional term in the square brackets is introduced to correct the effect of the curvature ($N = 2$, see Methods section 4.14, and Supplementary Notes 4 and 5). The phase velocity of the circumferential guided waves $c_c^p$ is determined by

$$c_c^p = \frac{\omega}{\mathrm{Re}(k)} \quad (3)$$

where the relationship of $\omega$ and $k$ implicitly depends on viscoelastic parameters of the artery $\alpha_c$, $\gamma$, $g$, $\tau$ (where $\alpha_c$ describe the arterial nonlinear elasticity in the circumferential direction), and wall thickness $h$ (see Eq. (15) in Methods section 4.14).

In the dispersion relations of guided axial and circumferential waves, we note the arterial stiffness and stress implicitly come into play through the parameters $\alpha_a$ and $\alpha_c$, respectively. To demonstrate, we reformulate $\alpha_a$ and $\alpha_c$ to decouple them as $\alpha_a = \sigma_a + \mu_{zr}\lambda_r$ and $\alpha_c = \sigma_c + \mu_{\theta r}\lambda_r$, where $\sigma_a$ and $\sigma_c$ denote axial and circumferential stresses, respectively; $\mu_{zr}$ and $\mu_{\theta r}$ denote tangent shear moduli at the stretched state; $\lambda_r$ is the stretch ratio in the radial direction (see details in Supplementary Note 6). The equations indicate the stiffness ($\mu_{zr}$ and $\mu_{\theta r}$) and stress ($\sigma_a$ and $\sigma_c$) jointly affect the parameters $\alpha_a$ and $\alpha_c$ and thus the dispersion relations. Interestingly, the stress $\sigma_a$ can be directly determined by using $\sigma_a = \alpha_a - \gamma$ (see Methods, section 4.16) once $\alpha_a$ and $\gamma$ have been obtained from the dispersion relations.

In experiment, the spatiotemporal data of the particle velocity are sampled with high frequencies in time and axial direction. Therefore, we can extract the full dispersion relation for the axial waves by Fourier transforming the spatiotemporal data to the wavenumber-frequency space. We find windowing the spatiotemporal field can effectively mitigate the influence of circumferential guided wave signals on the estimation of axial guided wave dispersion (see Methods section 4.6 and Supplementary Note 7). The dispersion relation for circumferential waves, however, can not be recovered due to the low sampling frequency (only one acquisition) in circumferential direction.

We then consider extracting arterial bidirectional stiffnesses $\alpha_a$ and $\alpha_c$, and bidirectional stresses $\sigma_a$ and $\sigma_c$ from the guided wave motions. Figure 4A and B depict our method. Briefly, we seek to fit the dispersion relations of axial guided waves to get $\alpha_a$ and $\gamma$. The nonlinear fitting was implemented by a genetic algorithm-aided method (see Methods, section 4.15). This fitting enables the measurement of the axial stress $\sigma_a = \alpha_a - \gamma$. In the circumferential direction, the circumferential stress $\sigma_c$ is accessible by measuring the blood pressure and calculating it using the Young-Laplace equation (*43*). The circumferential stiffness, $\alpha_c$, then can be recovered by $\alpha_c = \sigma_c + \gamma$.

To verify the inversion proposed here, we performed an *ex vivo* experiment on a porcine aorta. The estimation errors for both the axial stress and axial stiffness, compared to the uniaxial tensile test, were less than 10% (see Supplementary Note 8). Then we applied the inversion to the *in vivo* data for normal and craning postures. As shown in Figs. 4C and D, the axial dispersion significantly increases from the normal state to the craning state, with the variation in axial stress between the two postures measured at ~ 13 kPa. Existing methods for measuring arterial axial stress rely on known arterial constitutive model parameters and axial deformation (*34, 35*), which are indeed difficult to measure *in vivo*. Our method overcomes this limitation and enables the direct measurement of arterial axial stress through elastic waves, potentially enhancing the accuracy of axial stress measurement.

Figure 4E shows the *in vivo* results of bidirectional stresses in the young group. Significant difference was found between the axial and circumferential stresses both at diastole ($p < 0.001$) and systole ($p < 0.001$), where circumferential stress is higher than the axial ones. Figure 4F compares the bidirectional stiffnesses $\alpha_a$ and $\alpha_c$ in the young group. Significant difference was observed between $\alpha_a$ and $\alpha_c$ both at diastole ($p < 0.01$) and systole ($p < 0.001$), with circumferential stiffness being approximately 1.2 to 1.5 times higher than axial stiffness. Unlike bidirectional stiffnesses, bidirectional wave velocities in the young group were observed to be relatively similar, primarily due to the dominance of different modes in axial and circumferential guided waves.

Figure 4G shows representative dispersion curves of axial guided waves at diastole and systole among the three groups (inferred values of arterial mechanical parameters are listed in Table 2). Figure 4H and I compare the axial and circumferential stresses among the three groups, respectively. The circumferential stress increases with aging and hypertension, which is reasonable given the rise in blood pressure in the older and hypertensive groups. On the other hand, no significant differences were observed in axial stresses among the three groups, except that the systolic axial stress in the hypertensive group was significantly higher than that in the other two groups ($p < 0.01$). Interestingly, the axial stress keeps basically stable with aging and hypertension at diastole; this may be due to tissue-level responses (such as thickening and axial shortening) in order to restore wall axial stress towards homeostatic targets. Previous work estimated the axial stresses of CCAs in normotensive group; our results were consistent with their values (*34*). Figure 4J compares the axial and circumferential stiffness at diastole across the three groups, respectively. At diastole, axial stiffness showed no significant differences between the three groups, while circumferential stiffness in the older groups (both normotensive and hypertensive older groups; $p < 0.05$ and $p < 0.01$, respectively) was higher than that in the young group. This indicates that circumferential stiffness of arteries undergoes more pronounced changes with aging compared to axial stiffness. Our method to characterize bidirectional stresses and stiffnesses in arteries holds promise to provide a more comprehensive understanding of aging and hypertension.

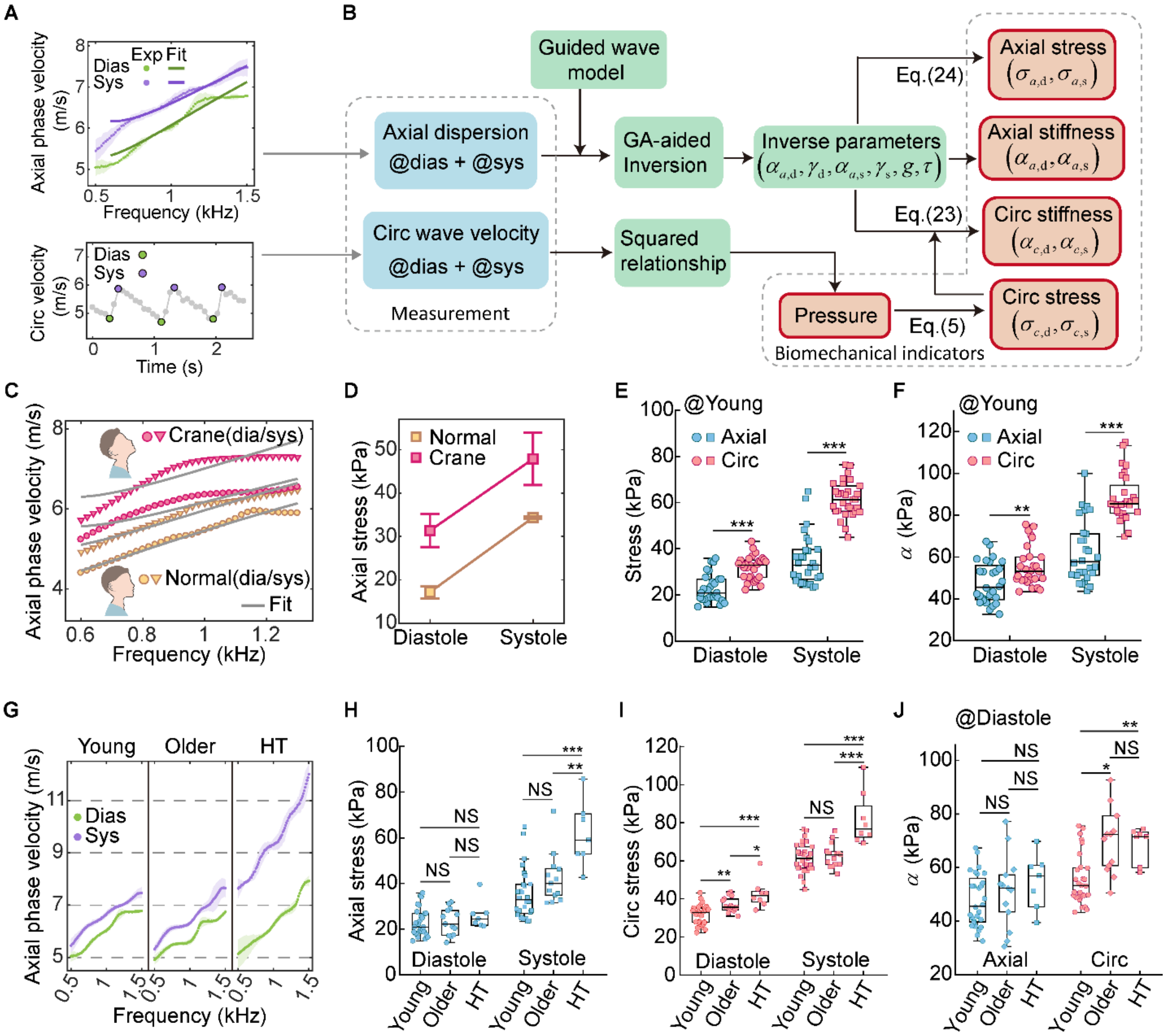

**Fig. 4. Characterization of** arterial **bidirectional stresses and bidirectional stiffnesses. (A)** Top: *In vivo* measurement of axial guided wave dispersion at diastole (representing the minimum wave dispersion within a single cardiac cycle) and systole (representing the maximum wave dispersion within a single cardiac cycle), along with their fitting curves. Bottom: *In vivo* measurement of circumferential wave velocities in cardiac cycles. **(B)** Flowchart to characterize arterial biomechanical indicators, including blood pressure, bidirectional stresses and bidirectional stiffnesses. **(C)** Dispersion curves for the two postures (normal and craning) and their fitting results. **(D)** Axial stresses in the two postures at diastole and systole. The axial stresses at diastole and systole in normal posture are 17.2±1.4 kPa, 34.4±0.2 kPa, respectively. The axial stresses at diastole and systole in craning posture are 31.3±3.8 kPa, 47.9±6.0 kPa, respectively. **(E)** Comparison of bidirectional stresses at diastole and systole, respectively, in the young group. **(F)** Comparison of bidirectional stiffnesses at diastole and systole, respectively, in the young group. **(G)** Measurement results of axial guided wave dispersion from three volunteers across the young, older, and hypertensive groups. **(H)** Comparison of axial stress at diastole and systole, respectively, among the three groups. **(I)** Comparison of circumferential stress at diastole and systole, respectively, among the three groups. **(J)** Comparison of axial stiffness and circumferential stiffness at diastole among the three groups, respectively.

**Table 2. Inferred values of the arterial mechanical parameters in the three groups**

| | $\alpha_{a,\mathrm{d}}$ (kPa) | $\alpha_{a,\mathrm{s}}$ (kPa) | $\alpha_{c,\mathrm{d}}$ (kPa) | $\alpha_{c,\mathrm{s}}$ (kPa) | $\gamma_{\mathrm{d}}$ (kPa) | $\gamma_{\mathrm{s}}$ (kPa) | $g$ | $\sigma_{a,\mathrm{d}}$ (kPa) | $\sigma_{a,\mathrm{s}}$ (kPa) | $\sigma_{c,\mathrm{d}}$ (kPa) | $\sigma_{c,\mathrm{s}}$ (kPa) |
|---|---|---|---|---|---|---|---|---|---|---|---|
| Young (*n*=30) | 47.8±9.8 | 62.3±14.8 | 55.6±9.5 | 88.1±11.3 | 24.1±6.6 | 26.8±6.3 | 0.77±0.08 | 22.8±6.1 | 35.2±11.0 | 31.7±5.1 | 61.5±7.6 |
| Older (*n*=14) | 51.1±13.8 | 73.3±17.8 | 63.8±10.1 | 94.7±12.6 | 27.2±8.3 | 31.4±9.2 | 0.76±0.11 | 23.1±5.9 | 42.1±11.1 | 36.8±3.8 | 62.7±6.1 |
| HT (*n*=8) | 54.9±10.3 | 94.0±15.6 | 68.8±6.7 | 113.8±11.6 | 26.5±6.8 | 32.1±9.5 | 0.80±0.06 | 26.2±6.3 | 61.7±14.2 | 42.6±7.3 | 81.7±13.8 |

Note: The viscous parameter $\tau$ ranges from approximately $3\times10^{-5}$ to $8\times10^{-5}$ s in the three groups.

## 3. Discussion

Arterial anisotropic stiffness, local blood pressure, and mechanical stresses within the arterial wall are intertwined indicators, crucial for understanding cardiovascular physiology and pathology (*1, 3*). The interplay among these parameters not only governs arterial wall mechanics but also directly impacts the progression of cardiovascular diseases. Historically, *in vivo* attempts have been made to investigate arterial axial stiffness (*19-21, 27*), circumferential stiffness (*29*), bidirectional stiffnesses (*44*), local blood pressure (*39, 40*), and axial stress (*34, 35*). Previous techniques typically assess these biomechanical indicators separately and are unable to capture their dynamic interrelations, which may ascribe to the temporal and spatial variabilities of arterial mechanics. In this study, we overcome these limitations by developing an ultrasound elastography method that utilizes programmed ARF to excite both the anterior and posterior walls of the artery. This approach allows for the simultaneous evaluation of axial and circumferential guided waves by imaging the tissue particle vibration along the arterial long-axis. By combining advanced acoustoelastic modeling and image processing techniques, our method enables the synchronous assessment of anisotropic stiffness, local blood pressure, and mechanical stresses in arteries in a single measurement. The clinical utility of our method has been demonstrated via *in vivo* experiments on healthy young, normotensive older, and hypertensive older volunteers. The results and findings highlight the capability of our method for providing a more comprehensive understanding of age- and disease-related changes in arterial biomechanics, enabling more accurate risk assessment, early disease detection, and personalized treatment strategies for patients with cardiovascular conditions.

The changes in arterial anisotropic stiffness are closely related to the distribution of collagen fibers (*45*). While many *ex vivo* experiments have characterized arterial anisotropy (*46-50*), discrepancies between *in vivo* and *ex vivo* conditions demand the measurement of artery anisotropy *in vivo*. The dynamic variations in arterial stiffness throughout the cardiac cycle and across different spatial locations pose significant challenges for *in vivo* characterization. By simultaneously capturing bidirectional guided waves, our method enables the assessment of anisotropic stiffness at specific cardiac phase and arterial location. Experimental results indicate that the circumferential stiffness of the right CCA is approximately 1.3 times higher than its axial stiffness. This method holds the potential to create high spatial and temporal resolution maps of arterial anisotropy by tuning wave frequencies, offering a powerful tool for the early detection of cardiovascular diseases.

Arterial stress, both axial and circumferential, plays a critical role in determining the mechanical behavior of the arterial wall and is closely linked to the progression of cardiovascular diseases (*33, 36*). Traditional methods for measuring axial stress often rely on the assumption of known constitutive parameters and deformation (*34, 35*), which is indeed difficult to achieve *in vivo*. In contrast, our method provides a direct means for measuring axial stress by analyzing the dispersion of axial guided waves without invoking prior knowledge of deformation or material properties. Our measurements show that the axial stress in the right CCA increases by approximately 13 kPa when transitioning from a forward gaze to a fully extended neck position. Notably, hypertension-induced remodeling of arterial walls is reflected in the elevated circumferential stress (a relative increase of 17% ~ 30% in cardiac cycles). Whereas, the axial stress exhibits relatively less changes (a relative increase of 0 ~ 25%) compared to the normotensive older individuals, suggesting a differential mechanical adaptation in response to chronic hypertension. The ability to assess both axial and circumferential stress in real-time during cardiac cycles opens up new possibilities for understanding the mechanobiology of arterial disease and could potentially impact the management of conditions like aortic dissection, aneurysm, and hypertension.

In addition to its implications for arterial biomechanics, our method enables non-invasive continuous blood pressure measurement, leveraging the linear correlation observed in this study between the square of circumferential wave velocities and blood pressure. Unlike conventional techniques such as applanation tonometry, our method eliminates the need for vascular compression, making it more promising for long-term blood pressure monitoring. Additionally, compared to the methods that estimate blood pressure based on arterial geometry measurement (*39, 40*), the present method is less affected by posture changes. According to our experiments, when the neck is fully extended, the diameter of the right common carotid artery reduces by approximately 0.2 mm (a ~3% decrease in the diameter change) compared to the forward gaze, whereas the circumferential wave velocities remain stable under varying neck postures (less than a 1% decrease in velocity change). This makes circumferential wave velocities as a highly promising index for continuous, long-term and reliable blood pressure monitoring, particularly when integrated with wearable ultrasonic devices (*39, 40, 51, 52*).

Beyond clinical applications, our method could be useful in broader interdisciplinary fields, including tissue engineering and soft robotics. The capability to simultaneously measure stiffness, stress, and pressure within soft tubular structures makes this method particularly relevant for evaluating artificial blood vessels (ABVs) used in biomedical engineering (*53*). By providing real-time feedback on the mechanical properties and stress distribution within ABVs, our approach could aid in optimizing their design and ensuring functional stability under physiological conditions. Additionally, the method offers a practical solution for monitoring soft robotic components, such as flexible tubing systems exposed to dynamic pressure environments (*54, 55*). Given the susceptibility of soft materials to environmental factors like temperature and pH, the capability to synchronously measure several key mechanical parameters *in situ* in a real-time manner could significantly advance their diverse applications.

## 4. Materials and Methods

### 4.1 Study design

We recruited 44 normotensive volunteers and 8 hypertensive patients. Participants with systolic blood pressure over 140 mmHg were grouped into hypertension (HT). Other healthy volunteers were divided into two groups, including young group ($n$ = 30, 20±2 years old), and older group ($n$ = 14, 51±8 years old). The older group has similar age range to the HT group ($n$ = 8, 55±5 years old). Table S1 lists the baseline characteristics of all the participants. All the participants signed an informed consent form before the experiments. Participants were asked to sit quietly for 10 minutes before the test in a sitting position.

At the beginning and end of experiments, brachial blood pressure was measured with an electronic sphygmomanometer (Upper Arm Blood Pressure Monitor HEM-7200, Omron, Japan). Arterial ultrasound elastography was conducted with an ultrasound system capable of generating programmed acoustic radiation force and performing ultrafast imaging (Verasonics Vantage 64LE System, Verasonics, USA). In the measurement, a L9-4 (central frequency 7 MHz, element number 124) linear array transducer (JiaRui Electronics, China) was used. Participants sat upright with a forward gaze during imaging. The probe was placed along the long-axis section of the right common carotid artery (CCA) with ultrasound gel between the probe and the skin to avoid the compression of the CCA. Electrocardiogram (ECG) signal was monitored (ADInstruments, New

Zealand) simultaneously during ultrasound elastography. After the elastography, a commercial ultrasound imaging system (Clover 60, Wisonic, China) equipped with a L15-4 (central frequency 9.5 MHz, element number 124) linear array transducer was used to measure the wall thickness precisely. The ultrasound probe was placed at the same position of the CCA, and the arterial walls in long-axis view were imaged. Afterwards, a pressure tonometer system (SPT-301, Millar Instruments, USA) relying on the principle of applanation tonometry (*38, 56*) was utilized to measure the blood pressure waveform of the CCA. The ECG signal was recorded simultaneously.

The protocol was approved by the institutional review board at Tsinghua University.

**4.2 Imaging protocol**

The ARF was imposed on the proximal artery (see Fig. 1B), and the guided wave propagating downstream was measured in this study. We defined a coordinate system $x$-$z$ in the imaging plane, where $x$ and $z$ denote vertical and lateral direction, respectively. The origin ($z = 0$) is located at the focus of ARF. Fig. 1c shows the imaging sequence. The ARF was generated by the focused ultrasound beams (voltage ~10 V). A single ARF (i.e., one push) lasted ~43 μs (300 cycles). Ten focused ARFs were imposed successively along the depth from ~1 mm above the anterior wall to ~1 mm below the posterior wall (F-number ~1.5). This sequential push (multiple ARFs) took in total ~0.5 ms and induced elastic Cherenkov effect (*57*) in arterial walls. The ultrasound system rested a duration of ~0.3 ms after excitation due to the voltage conversion of the system. Then the plane wave imaging started at a pulse repetition frequency of 10 kHz (sampling period 0.1 ms). The transducer transmitted and received unfocused ultrasound beams each time to reconstruct one frame. We designed an acquisition of 40 frames for plane wave imaging (~4 ms). In phase/quadrature (I/Q) data at both anterior and posterior walls were acquired. Although the guided wave was measured preferentially on the anterior wall in this study, signals on the posterior wall was shown to be vital for older and hypertensive volunteers. The tissue particle velocity $v_x\left(x,z,t\right)$ was estimated offline using the two-dimensional cross-correlator (*58*) with a kernel size of 5×2 (0.275 mm in $z$, 0.2 ms in $t$). A spatial filter (mean filter) with a kernel size of 4×4 (0.22 mm in $z$, 0.44 mm in $x$) was employed to smooth particle velocity field. This single measurement (guided waves excitation and detection) took ~5 ms. We repeated the measurement described above for 36 times to continuously measure arterial biomechanics through cardiac cycles; the system rested 65 ms between two adjacent measurements. The 36 measurements totally took ~2.5 s, covering approximately 3 cardiac cycles (see Fig. 1C).

The safety of the imaging method has been confirmed in our previous work (*20*).

### 4.3 Measurement of arterial geometry

To get arterial diameter waveform, the first frame of each single measurement was extracted and as a result we obtained 36 frames in total. B-lines at the same particle location from 36 frames were extracted and combined as an M-mode image. Then we measured inner arterial diameter from the image (*59*). By correcting the value of diameter at the first line of M-mode image, the diameter waveform at the centerline of the wall was finally obtained (see details in Supplementary Fig. S6). The wall thickness was measured at the anterior wall from the B-mode image, as a whole of three layers (including intima, media and adventitial layer). The wall thickness was identified according to literature (*60*). Limited by the spatial resolution of ultrasound (~0.15 mm), we only measure the wall thickness at end-diastole for further analysis.

### 4.4 Measurement of axial wave velocity

The axial wave velocity was measured by the time-to-peak method (TTP) (*61*). Firstly, two windowing boundaries, i.e. the lower-cut and upper-cut were added on the spatiotemporal map, aiming to remove circumferential and other noise signals. Then a region with higher signal-to-noise ratio (SNR) of axial wave was selected from the map. On this local map, (negative) peak point of tissue particle velocity versus time was recognized along each particle location. Linear regression of the relationship between the peak velocity arrival time and the distance along the anterior wall yielded the slope as the axial wave velocity. The coefficient of determination $r^2$ was used to evaluate the goodness of fit. The result of this measurement is shown in Fig.2a.

A comparison of the TTP method with another mostly-used method - Radon transformation (*62*) was conducted (see Supplementary Fig. S7). Both methods show good results of estimation at diastolic state, while at the early systolic state, the TTP method provides a better fitting than Radon transformation. Therefore, the TTP method provides a robust estimation of axial wave velocity during cardiac cycles.

### 4.5 Measurement of circumferential wave velocity

To obtain the arrival time of circumferential guided waves, the particle velocity versus time at the origin of the length of the spatiotemporal map was drawn out, which theoretically coincided with the motivation location. The velocity curve was smoothed by moving average with a kernel size of 5, and the arrival time $t_1$ was obtained by searching the first negative peak (generally in the range of 1 ~ 2 ms). The circumferential wave velocity was then calculated by

$$c_c = \frac{\pi r_c}{t_1 + t_c} \tag{4}$$

where $r_c$ is the radius of the artery. $t_c$ (= 0.8 ms) is the compensation time. In the experiment, the anterior and posterior walls were excited sequentially, causing certain uncertainties in the compensation time when calculating $c_c$ using signals from each wall. However, using a compensation time of 0.8 ms for both walls yields an estimation error of less than 6% (see Supplementary Note 9).

### 4.6 Extraction of phase velocity of axial guided waves

To extract the dispersion curve describing the variation of phase velocity with frequency, the spatiotemporal map was firstly preprocessed by applying windowing boundaries and removing mean value. Two windowing boundaries (*22*), including lower and upper cut, are set to ensure that signals from axial guided wave are captured and circumferential guided wave are removed. Then two-dimensional fast Fourier transformation (2D-FFT) was applied to the preprocessed map to get the *k*-space. By searching peaks at each frequency we obtained a *f*-*k* curve (*f* denotes frequency and *k* denotes wavenumber), and the phase velocity was derived by $c_p = f / k$ (see Supplementary Note 7).

### 4.7 Calibration of the carotid blood pressure

The carotid pressure waveform obtained by applanation tonometry was calibrated by the Kelly and Fitchet method (*38, 56*). This calibration is based on the principle that the diastolic (DBP) and mean blood pressure (MBP) is constant through the large artery tree. Brachial diastolic and systolic blood pressure (SBP) were acquired by the sphygmomanometer. MBP was calculated by MBP=SBP/3+2×DBP/3 (*63*).

Based on the linear relationship between the square of axial / circumferential wave velocity and the BP, the wave velocity could be utilized to measure the carotid pressure waveform, and the calibration of BP was conducted similar to the tonometry.

### 4.8 Measurement of circumferential stress of arteries

After measurements of blood pressure $P$ and arterial radius $r_c$ and wall thickness $h$, the circumferential stress of the arteries can be determined by the Young-Laplace equation (*43*),

$$\sigma_c = \frac{Pr_c}{h} \tag{5}$$

### 4.9 Phantom preparation and experiments

A homogeneous polyvinyl alcohol cryogel (PVA) tube was prepared. The PVA solution consisted of (by weight) 85% distilled water, 14% PVA, and 1% Sigmacell cellulose. The PVA and cellulose were mixed into 85°C water and stirred until fully dissolved. Then the PVA solution was poured into a tube mold, and underwent solidification and polymerization by three freezing-thawing cycles. A freeze-thaw cycle lasted 48 hours, with -20°C freezing and 20°C thawing. The inner diameter and wall thickness of the tube phantom were 6.5 and 1.5mm, respectively.

After the tube phantom was prepared, it was immersed in water and driven by the designed ARFs. The imaging settings were basically consistent with those of *in vivo* experiments, except for the following modifications: a single imaging sequence was applied to the phantom, the voltage of the focused ultrasound beams was increased to ~20 V to enhance SNR, and the plane wave imaging duration was set to ~7 ms. To characterize the elastic properties of the phantom in the quasi-static state, tensile tests (ElectroForce 3200, TA Instruments, USA) were also performed on the phantom samples (see Supplementary Note 2).

### 4.10 Ex vivo experiments

To verify the inversion method of measuring axial stress, an *ex vivo* experiment was performed. A segment of porcine thoracic aorta was obtained from a freshly slaughtered animal within 12 hours. Before the experiment, the aorta was cut into a length of ~19 cm. The wall thickness of the aorta was ~ 3.5 mm and the inner radius of the aorta was ~6.5 mm. In the ultrasound experiment, the aorta was cannulated at both ends and underwent different longitudinal pre-stretches, including $\lambda_z = 1$ (stress-free), 1.23 and 1.31. A single imaging sequence was performed on the phantom. The voltage of the focused ultrasound beams was set to ~20 V to get a higher SNR. The duration of the plane wave imaging was ~7 ms. To characterize the hyperelastic properties of the aorta, tensile tests (ElectroForce 3200, TA Instruments, USA) were also performed on the aorta samples (see Supplementary Note 8).

### 4.11 Finite element analysis

Finite element analysis was performed using Abaqus/CAE 6.14 (Dassault Systemes, USA). In order to simulate nonlinear viscoelasticity of arteries, we adopted the HGO model (*64*) and first-order Prony series model. The arterial wall was immersed in water. Firstly, the tube underwent a quasi-static deformation caused by the inner blood pressure and axial pre-stretch. Then a constant body force with a Gaussian distribution simulating the ARF was applied on both the anterior and posterior walls of the tube, which induced guided waves propagating along both the axial and circumferential directions (see Supplementary Fig. S8). Approximately 200,000 solid elements (C3D8R) were used to model the arterial wall and about 300,000 acoustic elements (AC3D8) applied to discrete the water. Convergence of the simulation was carefully examined by comparing the computational results with those given by a refining mesh and a smaller time step.

### 4.12 Statistical analysis

Statistical analysis was preformed using MATLAB R2016b (MathWorks Inc., USA). All experimental data were expressed as mean ± standard deviation (s.d.). Pearson's correlation coefficient $r$ was adopted to evaluate the correlation of the square of wave velocities with blood pressure. Bland-Altman analysis was used to compare two methods for measuring blood pressure: the elastography method (which utilizes circumferential wave velocities), and applanation tonometry. The unpaired $t$-test was performed to compare bidirectional wave velocities, mechanical stiffnesses and stresses. A $p$-value of 0.05 was adopted to indicate statistical significance. Results with $0.01 \leq p \leq 0.05$ were marked with one asterisk (*); those with $0.001 \leq p \leq 0.01$ were marked with two asterisks (**); and those with $p < 0.001$ were marked with three asterisks (***).

### 4.13 Constitutive modelling of nonlinear viscoelastic arteries

The artery was modelled as a thin-walled tube subject to inner blood pressure $P$ and axial pre-stretch. A cylindrical system was built for the artery, with $\theta$, $r$, $z$ representing the circumferential, radial, and axial directions, respectively. $\lambda_\theta$, $\lambda_r$ and $\lambda_z$ denote the stretch ratio in circumferential, radial and axial direction, respectively. The incompressible condition yields $\lambda_\theta \lambda_r \lambda_z = 1$.

Fung's quasi-linear viscoelastic model was adopted to describe arterial nonlinear viscoelasticity (*65*). The Cauchy stress $\boldsymbol{\sigma}$ is defined as

$$\boldsymbol{\sigma} = -q\boldsymbol{I} + \int_0^t \Xi(t-s) \cdot \frac{\partial \boldsymbol{\sigma}_D^e(s)}{\partial s} \mathrm{d}s \tag{6}$$

where $q$ denotes the isochoric part of the stress, and the second term denotes the deviatoric part. $\Xi(t)$ is the relaxation function of the viscoelastic material. We employed the first-order Prony series model to represent arterial viscoelasticity. In comparison to the widely used fractional derivative model for biological tissues (*66*), this model can effectively captures the dispersion within the limited frequency bandwidth (e.g., 0.5 - 1.5 kHz) measured in the experiment (see details in supplementary Note 10). Thus we have

$$\Xi(t) = 1 - g\left[1 - \exp(-t/\tau)\right] \tag{7}$$

where $g$ and $\tau$ denote the relaxation amplitude and characteristic relaxation time, respectively. $\boldsymbol{\sigma}_D^e$ in Eq. (6) denotes the deviatoric part of elastic stress $\boldsymbol{\sigma}^e$. The elastic stress is related to the strain energy function $W$ and deformation tensor $\boldsymbol{F} = \mathrm{diag}(\lambda_\theta, \lambda_r, \lambda_z)$, according to $\boldsymbol{\sigma}^e = (\partial W / \partial \boldsymbol{F})\boldsymbol{F}^{\mathrm{T}}$. We adopted the HGO model to describe arterial hyperelasticity (*64*),

$$W = \frac{\mu}{2}(I_1 - 3) + \frac{k_1}{2k_2}\sum_{i=4,6}\left\{\exp\left[k_2\left[\kappa(I_1-3) + (1-3\kappa)(I_i - 1)\right]^2\right] - 1\right\} \tag{8}$$

where $\mu$ and $k_1$ are the initial shear modulus of elastin and collagen fibers, respectively. $k_2$ denotes the nonlinear stiffening. $\kappa$ represents the scattering of fibers (between 0 and 1/3, whereas 1/3 corresponds to the isotropic material). The model consists of two symmetrically distributed fibers, with vectors $\boldsymbol{M}_1 = (\cos\varphi, 0, \sin\varphi)^{\mathrm{T}}$ and $\boldsymbol{M}_2 = (-\cos\varphi, 0, \sin\varphi)^{\mathrm{T}}$, where $\varphi$ denotes the angle between the fiber orientation and the circumferential direction (see Supplementary Fig. S8). Invariants $I_1$, $I_4$ and $I_6$ are defined as $I_1 = \lambda_\theta^2 + \lambda_z^2 + \lambda_\theta^{-2}\lambda_z^{-2}$, $I_{4,6} = \lambda_\theta^2\cos^2\varphi + \lambda_z^2\sin^2\varphi$.

### 4.14 Guided axial and circumferential wave model for pre-stressed viscoelastic arteries

Imagine unrolling a cylindrical tube along its axial direction and unfolding it into a flat plate (see Supplementary Fig. S9). The dispersion relation for the guided waves propagating on this flat plate (Lamb wave) is given as follows. A Cartesian coordinate system was established on the plate, where $x_r$, $x_\theta$, and $x_z$ corresponds to the radial, circumferential, and axial directions of the artery, respectively. Consider that this viscoelastic pre-stressed flat plate is immersed in an inviscid fluid, with guided waves propagating along the axial direction. The excitation of ARF leads to the dominance of the antisymmetric mode of Lamb waves (*19*). The dispersion relation is given by (see derivation in Supplementary Note 11):

$$\begin{aligned}&\left(1+s_{2a}{}^{2}\right)\cdot\left(-\rho\frac{\omega^2}{k^2}s_{1a}+C_{1a}s_{1a}-C_{2a}s_{1a}{}^{3}\right)\cdot\tanh\left(s_{1a}kh/2\right)\\&-\left(1+s_{1a}{}^{2}\right)\cdot\left(-\rho\frac{\omega^2}{k^2}s_{2a}+C_{1a}s_{2a}-C_{2a}s_{2a}{}^{3}\right)\cdot\tanh\left(s_{2a}kh/2\right)+\left(s_{1a}{}^{2}-s_{2a}{}^{2}\right)\frac{\rho^f}{\xi}\frac{\omega^2}{k^2}=0\end{aligned} \tag{9}$$

where $h$ is the wall thickness. $s_{1a}$ and $s_{2a}$ are two roots solved by the quartic equation

$$\left(\gamma+\frac{2.5\alpha_a+\gamma}{3}\Omega\right)s^4+\left[\rho\frac{\omega^2}{k^2}-8G\alpha_a+\frac{\gamma-2\alpha_a}{3}\Omega\right]s^2+\alpha_a+\frac{2.5\alpha_a+\gamma}{3}\Omega-\rho\frac{\omega^2}{k^2}=0 \tag{10}$$

and

$$C_{1a}=8G\alpha_a+\gamma+1.5\Omega\alpha_a \tag{11}$$

$$C_{2a}=\gamma+\frac{2.5\alpha_a+\gamma}{3}\Omega \tag{12}$$

$$\xi^2=1-\frac{\omega^2}{k^2}\frac{1}{c_f{}^2} \tag{13}$$

$$G=\left(1-\frac{g}{1+i\omega\tau}\right)/\left(1-g\right),\ \Omega=\left(\frac{g\omega^2\tau^2}{1+\omega^2\tau^2}+i\frac{g\omega\tau}{1+\omega^2\tau^2}\right)/\left(1-g\right) \tag{14}$$

where incremental parameters $\alpha_a$ , $\gamma$ are defined as $\alpha_a=\mathcal{A}_{0zrzr}$ , $\gamma=\mathcal{A}_{0rzrz}=\mathcal{A}_{0r\theta r\theta}$ . $\mathcal{A}_{0jikl}=\left(\partial^2W/\partial F_{ip}\partial F_{lq}\right)F_{jp}F_{kq}$ $(j,i,k,l,p,q\in(r,\theta,z))$ is the fourth-order Eulerian elasticity tensor (see Supplementary Note 12 for explicit forms). $W$ denotes the strain energy function of arteries. $g$ and $\tau$ are viscous parameters. $i$ in Eq. (14) denotes the imaginary unit. The density of the arterial wall is $\rho$ = 1000 kg/m$^3$. The bulk modulus of the blood is $\kappa_p$ = 2.2 GPa. The density of the blood is $\rho^f$ = 1000 kg/m$^3$. The speed of sound in the blood is $c_f=\sqrt{\kappa_p/\rho^f}$. $\omega$ $(=2\pi f)$ denotes the angular frequency. $k$ denotes the wavenumber.

The dispersion of axial guided waves in a tube differs from that of Lamb waves (*24*), which reflects the curvature effect on wave dispersion. We further adopted a simplified model proposed by Li and Rose (*67*), which treats axial guided waves in a tube as Lamb waves in an unwrapped plate and imposes the periodic boundary condition on the circumferential direction of the plate. As a result, the phase velocity of axial guided waves, $c_{\mathrm{a}}^{p}$, can be calculated by Eq. (2), where $N$ (= 0, 1, 2,…) denotes the number of periodic waves in the circumferential direction. Indeed, the excitation of ARF beam leads to multiple modes in low frequencies (e.g. < 0.5 kHz) (*22*), and the ARF beam shape and location affects the dominant modes (*68, 69*). In this study, by sequentially focusing the ARF from the anterior to the posterior wall, we found that the dominant mode of the axial guided wave in high frequencies (e.g. > 0.6 kHz) corresponds to $N$ = 2 (see details in Supplementary Note 4).

As for the circumferential guided waves, both previous studies (*70*) and our FEA results (see Supplementary Note 5) have verified that the dispersion curve of the circumferential guided waves can be well approximated by the Lamb wave. Therefore, the dispersion of the circumferential guided waves is approximated by the antisymmetric mode of the Lamb wave as (see derivation in Supplementary Note 11):

$$\begin{aligned}&\left(1+{s_{2c}}^{2}\right)\cdot\left(-\rho\frac{\omega^{2}}{k^{2}}s_{1c}+C_{1c}s_{1c}-C_{2c}{s_{1c}}^{3}\right)\cdot\tanh\left(s_{1c}kh/2\right)\\&-\left(1+{s_{1c}}^{2}\right)\cdot\left(-\rho\frac{\omega^{2}}{k^{2}}s_{2c}+C_{1c}s_{2c}-C_{2c}{s_{2c}}^{3}\right)\cdot\tanh\left(s_{2c}kh/2\right)+\left({s_{1c}}^{2}-{s_{2c}}^{2}\right)\frac{\rho^{f}}{\xi}\frac{\omega^{2}}{k^{2}}=0\end{aligned}\tag{15}$$

where $s_{1\mathrm{c}}$ and $s_{2\mathrm{c}}$ are two roots solved by the quartic equation

$$\left(\gamma+\frac{1.7\alpha_{c}+\gamma}{3}\Omega\right)s^{4}+\left[\rho\frac{\omega^{2}}{k^{2}}-8G\alpha_{c}+\left(\frac{1}{3}\gamma-0.1\alpha_{c}\right)\Omega\right]s^{2}+\alpha_{c}+\frac{1.7\alpha_{c}+\gamma}{3}\Omega-\rho\frac{\omega^{2}}{k^{2}}=0\tag{16}$$

and

$$C_{1c}=8G\alpha_{c}+\gamma+0.7\Omega\alpha_{c}\tag{17}$$

$$C_{2c}=\gamma+\frac{1.7\alpha_{c}+\gamma}{3}\Omega\tag{18}$$

where incremental parameter $\alpha_c$ is defined as $\alpha_{c}=\mathcal{A}_{0\theta r\theta r}$. Eq. (15) determines the relationship between $\omega$ and $k$, and the circumferential phase velocity can be calculated by $c_{\mathrm{c}}^{p}=\omega/\mathrm{Re}(k)$.

## 4.15 Genetic algorithm-aided inversion to infer mechanical parameters of arteries

To infer arterial axial stiffness and stress from axial guided waves in arteries, we proposed a genetic algorithm-aided inverse method. The dispersion curve of axial guided waves at diastole (minimal pressure) and systole (peak pressure) were marked as $\left(f_{i,\mathrm{d}}, c_{i,\mathrm{d}}^{p}\right)$ and $\left(f_{i,\mathrm{s}}, c_{i,\mathrm{s}}^{p}\right)$, respectively ($i$ = 1, 2, …$n$, where $n$ denotes the amount of data points. The frequency range for diastolic data is 0.6 - 1.5 kHz, and 0.7 - 1.5 kHz for systolic curve, thus $n$ is approximately 45 and 40 for diastolic and systolic data, respectively). The arterial radius at diastolic and systolic states are $r_{c,\mathrm{d}}$ and $r_{c,\mathrm{s}}$, respectively. According to Eq. (9), two elastic incremental parameters $\alpha_a$ and $\gamma$, and two viscous parameters $g$ and $\tau$ are to be optimized from the experimental dispersion data. Assuming that the viscous parameters $g$ and $\tau$ remain relatively constant, while the incremental parameters $\alpha_a$ and $\gamma$ change significantly between diastolic (with subscript 'd') and systolic states (with subscript 's'), there are a total of six unknown parameters to be optimized in the two states, including $\alpha_{a,\mathrm{d}}$, $\gamma_{\mathrm{d}}$, $\alpha_{a,\mathrm{s}}$, $\gamma_{\mathrm{s}}$, $g$ and $\tau$. The optimization was achieved by the genetic algorithm (population size 50, crossover fraction 0.8, migration fraction 0.2). Compared to traditional iterative algorithms (e.g., Levenberg-Marquardt), it has advantages in global optimization for multi-parameter searches and does not rely on the gradient for its optimization direction. The best fit was obtained by minimizing the goodness-of-fit function (e.g. $\mathcal{F} < 0.1$ in practice),

$$\mathcal{F} = \left(\mathcal{F}_{\mathrm{d}} + \mathcal{F}_{\mathrm{s}}\right)/2 \tag{19}$$

where

$$\mathcal{F}_{\mathrm{d}} = \sqrt{\frac{\sum_{i=1}^{n}\left[c_{i,\mathrm{d}}^{p\,(\exp)} - c_{i}^{p(\mathrm{theo})}\left(\alpha_{a,\mathrm{d}}, \gamma_{\mathrm{d}}, g, \tau; f_{i,\mathrm{d}}, r_c, h\right)\right]^2}{n}}$$

$$\mathcal{F}_{\mathrm{s}} = \sqrt{\frac{\sum_{i=1}^{m}\left[c_{i,\mathrm{s}}^{p\,(\exp)} - c_{i}^{p(\mathrm{theo})}\left(\alpha_{a,\mathrm{s}}, \gamma_{\mathrm{s}}, g, \tau; f_{i,\mathrm{s}}, r_c, h\right)\right]^2}{m}} \tag{20}$$

where $c_i^{p(\exp)}$ and $c_i^{p(\mathrm{theo})}$ denote experimental and theoretically predicted axial phase velocity (Eq. (9)). The relationship between diastolic and systolic state provides constraints on material parameters, including $\alpha_{a,\mathrm{s}} > \alpha_{a,\mathrm{d}}$ and $\gamma_{\mathrm{s}} > \gamma_{\mathrm{d}}$. Parameter space was set as $20 \le \alpha_a \le 200\,\mathrm{kPa}$, $0 < \gamma/\alpha_a < 1$, $0.3 < g < 0.95$, $10^{-5} < \tau < 10^{-4}$ s (see details in Supplementary Note 13). As a result, six parameters were obtained by using the genetic algorithm-aided inversion, i.e., $\left(\alpha_{a,\mathrm{d}},\gamma_{\mathrm{d}},\alpha_{a,\mathrm{s}},\gamma_{\mathrm{s}},g,\tau\right)$.

Using dispersion data from both diastole and systole improves the stability of the inversion parameters compared to using data from a single state. Numerical examples were used to assess the stability of the inversion method, demonstrating that the estimated parameters exhibit acceptable stability; for example, the inversion error of $\alpha_a$ is within 5% (see Supplementary Note 14). The inversion efficiency was also evaluated by CPU runtime, requiring approximately 6 minutes on a standard laptop (see details in Supplementary Note 14). This also indicates the potential of the method for practical applications.

### 4.16 Relationship between atrial stress and incremental parameters

The general relationship between the stress and incremental parameters satisfies (*37, 71*)

$$\sigma_c-\sigma_r=\alpha_c-\gamma \tag{21}$$

and

$$\sigma_a-\sigma_r=\alpha_a-\gamma \tag{22}$$

where $\sigma_r$, $\sigma_a$, $\sigma_c$ represent the radial, axial, and circumferential stress, respectively. For arteries subjected to blood pressure and axial stretch, the radial stress is significantly smaller than both the axial and circumferential stress (*45*), hence we get the following approximate relationship

$$\sigma_c=\alpha_c-\gamma \tag{23}$$

and

$$\sigma_a=\alpha_a-\gamma \tag{24}$$

**Acknowledgments**

**Funding:**

Guo-Yang Li acknowledges the financial support from the National Natural Science Foundation of China (Grant No. 12472176), and the Fundamental Research Funds for the Central Universities, Peking University.

Xinyu Wang acknowledges support from the Haidian Innovation Transformation Fund (HDCXZHKC2021216).

Yanping Cao acknowledges support from the National Natural Science Foundation of China (Grants Nos. 11972206 and 11921002).

**Author contribution:**

**Conceptualization**: Y. J., G.-Y. L., and Y. C. **Methodology**: Y. J., G.-Y. L., K. H., Y. Z., X. W., and Y. C. **Investigation**: Y. J., G.-Y. L., K. H., S. M., Y. Z., M. J., Z. Z, X. W., and Y. C. **Visualization**: Y. J. **Funding acquisition:** G.-Y. L., X. W., and Y. C. **Project administration**: Y. C. **Supervision**: G.-Y. L., X. W., and Y. C. **Writing—original draft**: Y. J., G.-Y. L., and Y. C. **Writing—review & editing**: Y. J., G.-Y. L., K. H., X. W., and Y. C.

**Competing interests:** The authors declare that they have no competing interests.

**Data and materials availability:** All data needed to evaluate the conclusions in the paper are present in the paper and/or the Supplementary Materials.

**Supplementary Materials**

This PDF file includes:

Figs. S1 – S20

Supplementary Notes 1 – 15

Table S1 – S3

References

# Supplementary Materials for

## Simultaneous imaging of bidirectional guided waves enables synchronous probing of mechanical anisotropy, local blood pressure, and stress in arteries

Yuxuan Jiang *et al.*

Corresponding author: Guo-Yang Li, lgy@pku.edu.cn; Xinyu Wang, wangxinyu@bjmu.edu.cn;
Yanping Cao, caoyanping@tsinghua.edu.cn

**This PDF file includes:**

Figs. S1 to S20
Supplementary Notes 1 to 15
Tables S1 to S3
References

## Supplementary Figures

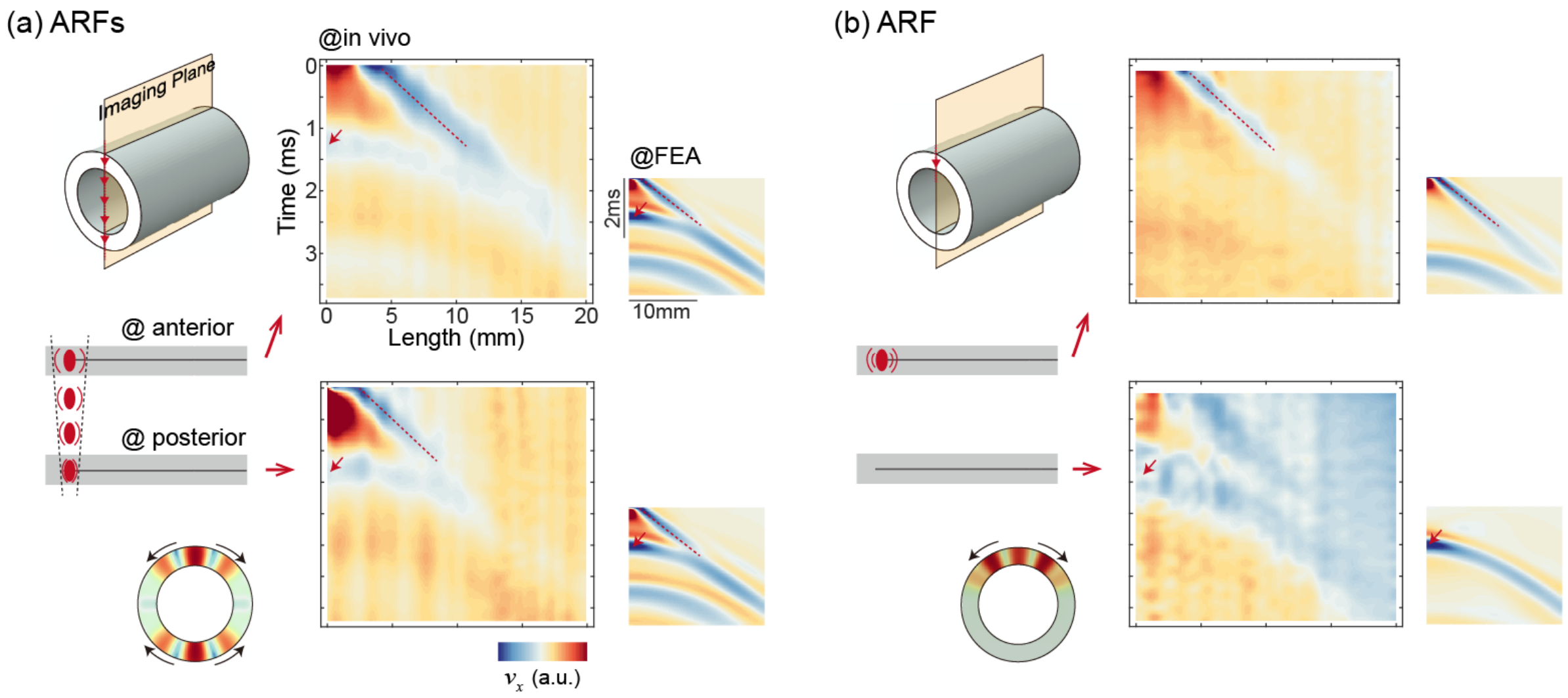

**Fig. S1**. An *in vivo* contrast experiment on a young healthy volunteer with (a) multiple ARFs on both anterior and posterior walls, (b) a single ARF on the anterior wall. The *in vivo* tissue particle velocity maps at end-diastole were extracted from both the anterior and posterior walls. FEA results are also shown for comparison with the *in vivo* results.

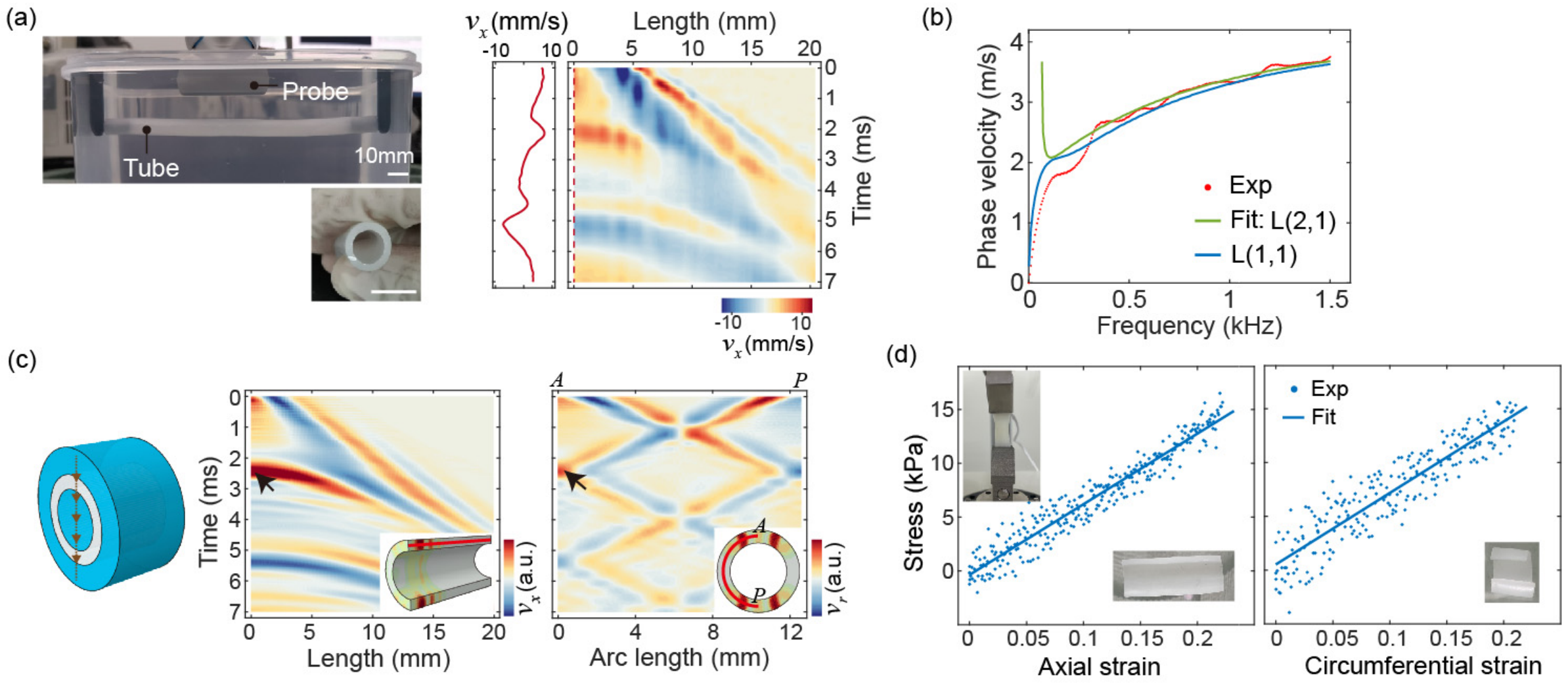

**Fig. S2**. Ultrasound elastography experiments on an artery phantom. (a) Experimental set-up and the particle velocity map measured from the anterior wall along the axial direction. (b) Mechanical characterization of the artery phantom by fitting the dispersion curve with the axial guided wave model. (c) Simulation results of FEA with the same geometry and material parameters as the phantom artery. (d) Mechanical characterization of the artery phantom by the uniaxial tension.

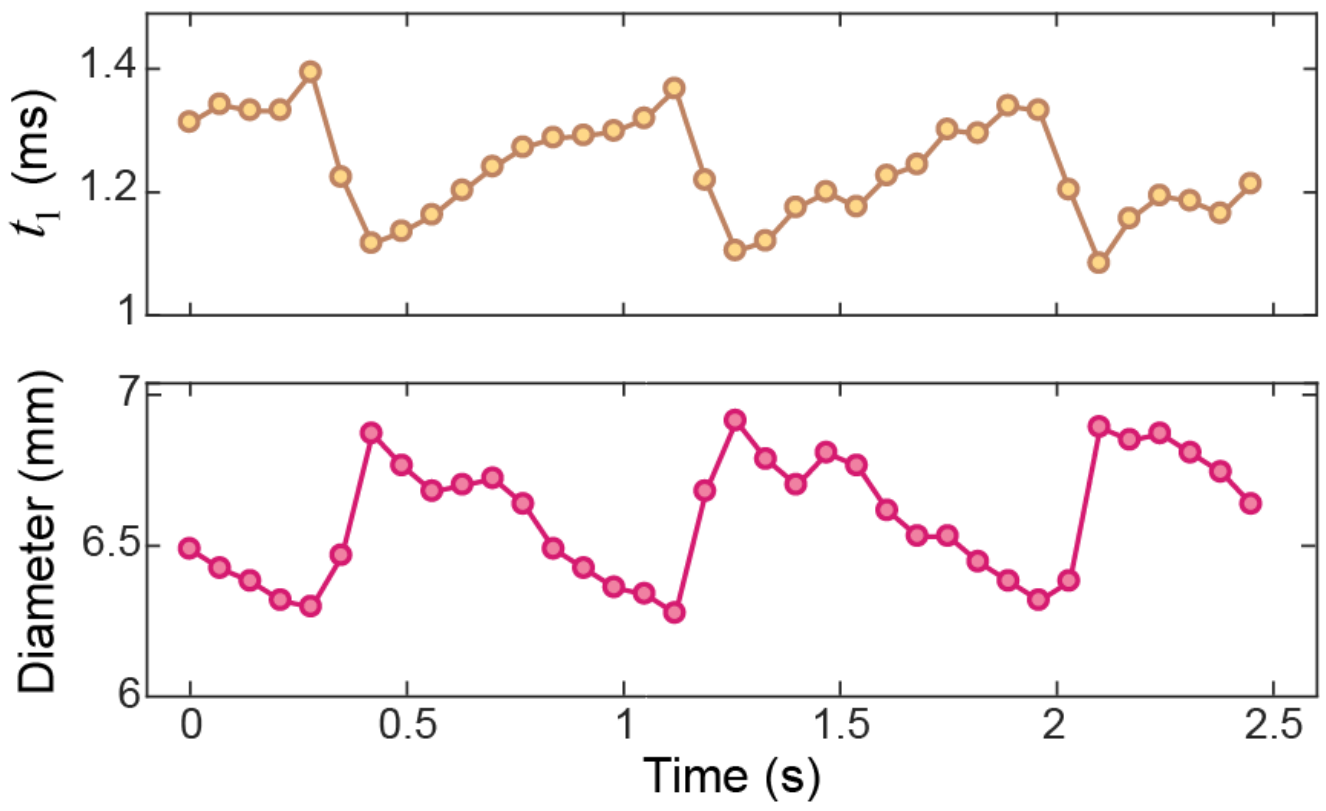

**Fig. S3**. Variation of the arterial diameter and circumferential time phase $t_1$ with respect to time in cardiac cycles.

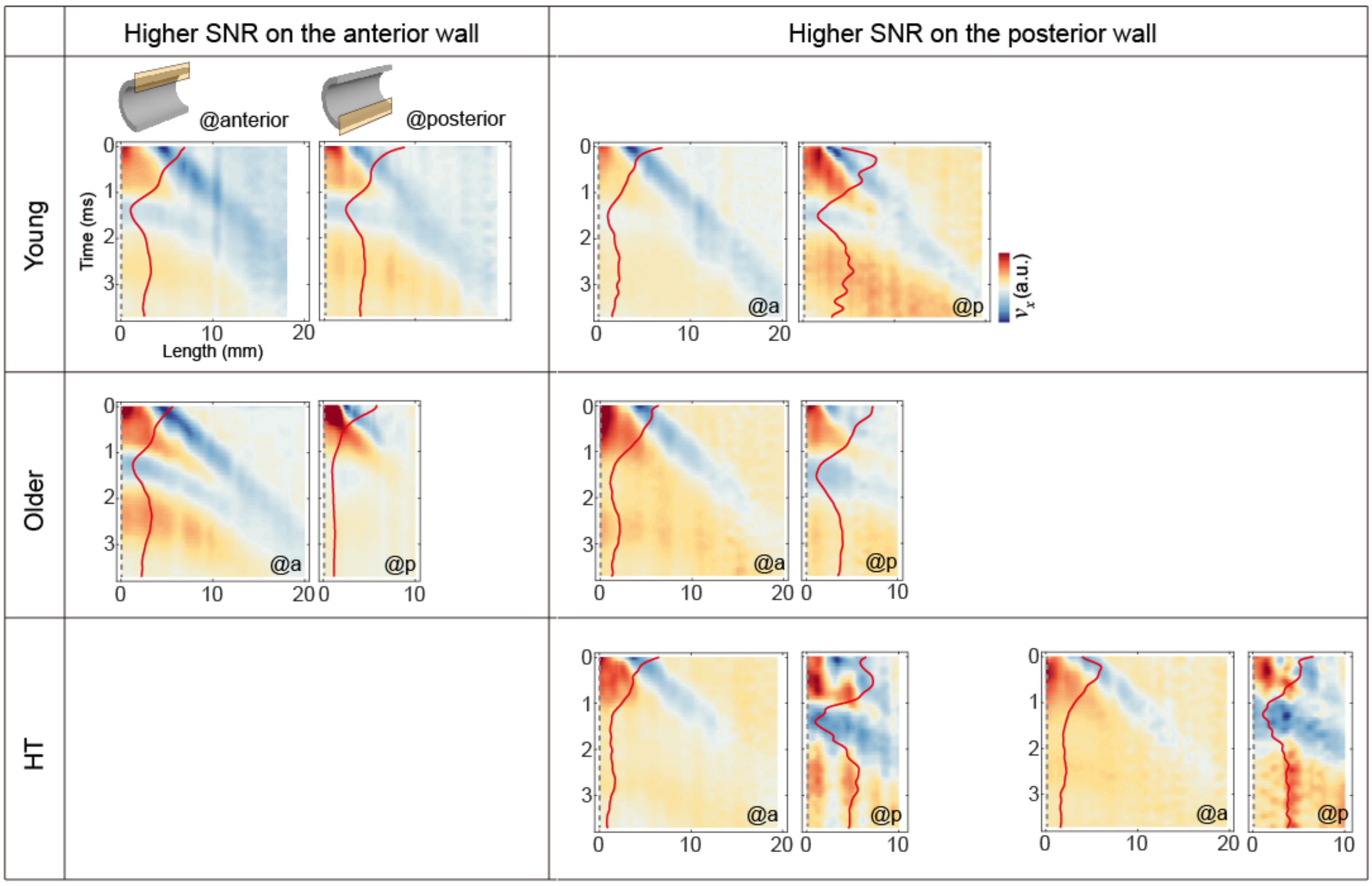

**Fig. S4**. Particle velocity maps extracted from both the anterior and posterior walls at end-diastole for the young, older and hypertension groups. For each group we show representative results from two individuals. These *in vivo* results can be divided into two categories by comparing SNR (signal-to-noise ratio) of peaks at time $t_1$ on the anterior and posterior walls.

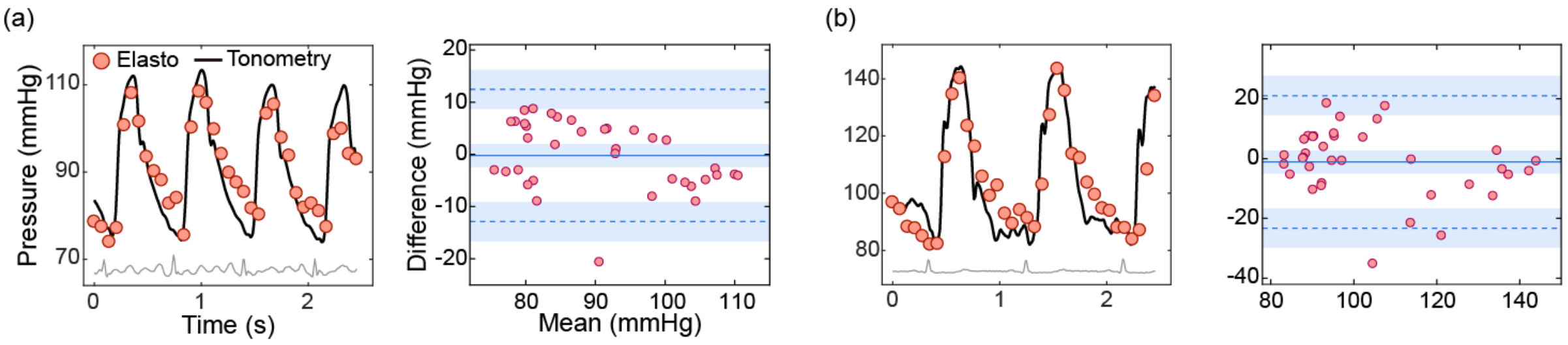

**Fig. S5**. Bland-Altman plot of the two blood pressure measurement methods, i.e., elastography and tonometry, for (a) an older participant (56 years old, male), and (b) a hypertensive participant (61 years old, female). The biases are: (1) older participant: -0.2 mmHg with LoA of -12.9 to 12.5 mmHg; (2) hypertension participant: -1.1 mmHg with LoA of -23.2 to 21.0 mmHg.

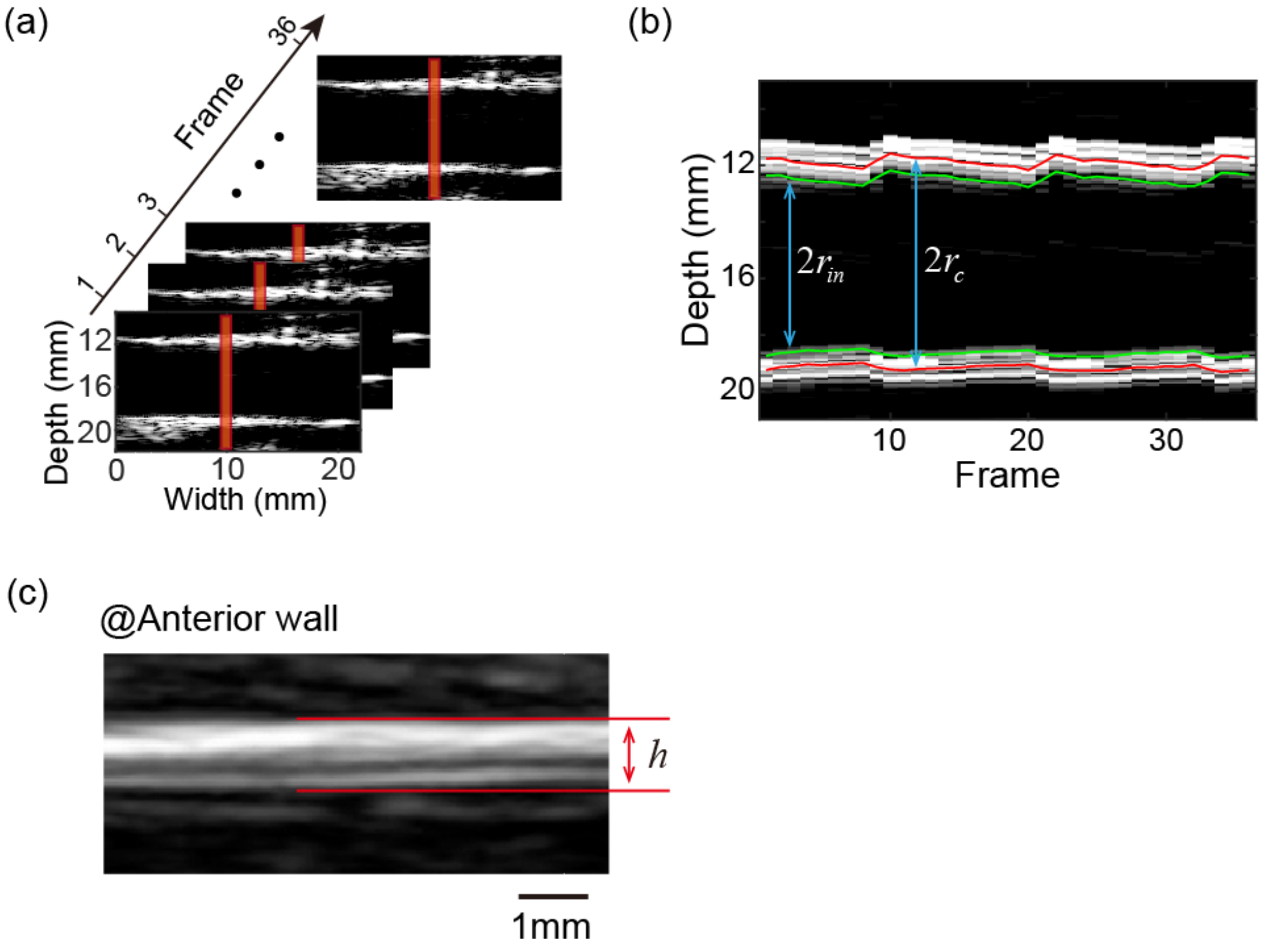

**Fig. S6**. Measurement of arterial diameter and wall thickness. (a) Extraction of M-mode image. In 36 repeated measurements, the first frame of the B-mode image obtained from each measurement was extracted. For each frame, the A-line image at the middle position (width @ 10mm) was selected. The 36 A-line images were then arranged in sequence to construct the M-mode image. (b) Measurement of the arterial diameter from the M-mode image. The inner radius $r_{in}$ is firstly detected; then the central radius can be obtained using $r_c = r_{in} + h/2$. (c) The wall thickness $h$ is measured at the anterior wall at end-diastole.

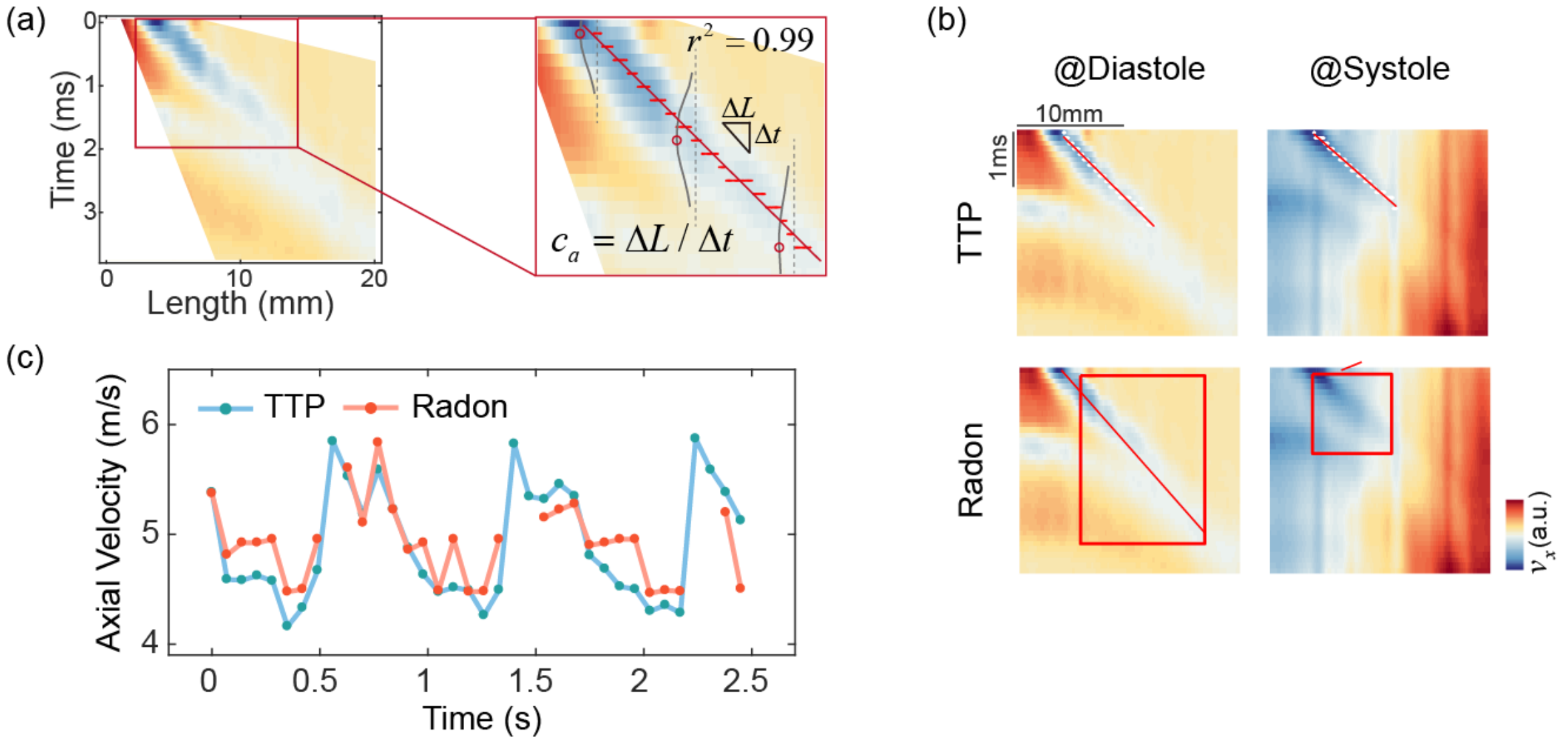

**Fig. S7**. Measurement of axial wave velocities. (a) The time-to-peak (TTP) method to measure axial wave velocities. (b) *In vivo* measurements using the TTP method and the Radon transformation method at diastole and systole. The Radon transformation method fails to fit the axial wave velocity at systole. (c) Comparison of the axial wave velocities measured by the TTP method and Radon transformation method. The wave velocity curve using Radon transformation method exhibits some discontinuities. In contrast, the TTP method is more robust and leads to a continuous wave velocity curve in cardiac cycles.

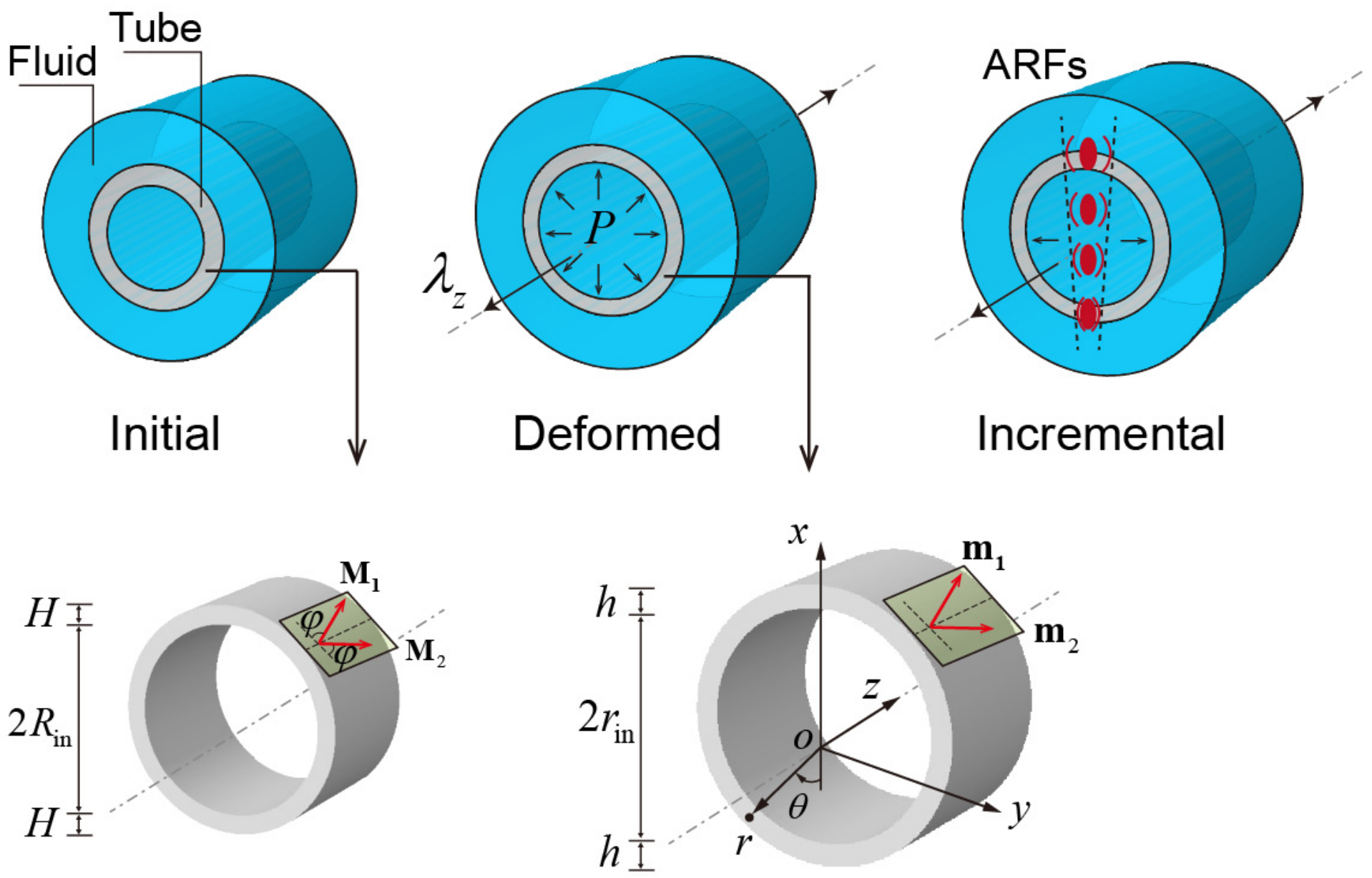

**Fig. S8**. Finite element model (FEM) to simulate arterial bidirectional guided waves. A fluid-immersed tube was constructed to simulate arterial conditions. Three mechanical configurations were built: the initial state (stress-free), the deformed state (artery subjected to blood pressure $P$ and axial stretch $\lambda_z$), and the incremental state (elastic waves excited by ARFs). The wall thickness and inner radius of the artery in the initial state are $H$ and $R_{in}$, respectively. The wall thickness and inner radius in the deformed state are $h$ and $r_{in}$, respectively. HGO model (*64*) was used in the FEM, with $\mathbf{M}_1$ and $\mathbf{M}_2$ representing two symmetric fiber orientations. These orientations deform into $\mathbf{m}_1$ and $\mathbf{m}_2$ under pressure and axial stretch.

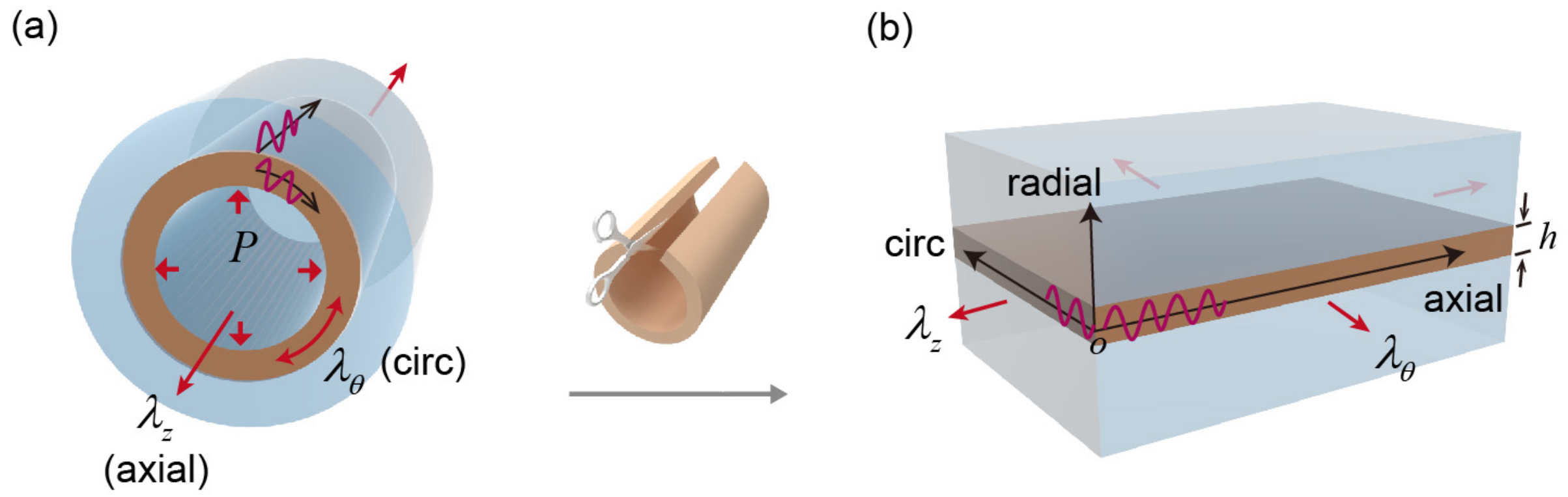

**Fig. S9**. Mechanical modeling of the arterial bidirectional guided waves. (a) The artery is modeled as a cylindrical tube immersed in fluid. The artery is subjected to blood pressure and axial stretch, with circumferential stretch ratio $\lambda_\theta$ and axial stretch ratio $\lambda_z$. (b) The tube is cut along the axial direction and unfolded into a flat configuration, with fluid on both the upper and lower sides. A Cartesian coordinate system is established on the plate, with the three directions referred to as axial ($x_z$), circumferential ($x_\theta$), and radial directions ($x_r$).

**Supplementary Note 1. An *in vivo* comparative experiment to verify the mechanical meaning of the time phase $t_1$**

An *in vivo* contrast experiment was conducted on a young healthy volunteer (27 years old, male) to verify the mechanical meaning of the time phase $t_1$. As depicted in Fig. S1a, when utilizing the ARFs excitation mode, guided waves were generated on both the anterior and posterior walls. Notably, the velocity maps extracted from the anterior and posterior walls exhibited the characteristics associated with the time phase $t_1$. In contrast, as shown in Fig. S1b, when using the ARF excitation mode, the guided waves were solely induced on the anterior wall. The time phase $t_1$ could only be detected on the map extracted from the posterior wall. The FEA results concurred with the observations made in the *in vivo* experiments. Therefore, the time phase $t_1$ denotes the time phase of the arrival of the circumferential guided wave originating from the opposite wall.

## Supplementary Note 2. Phantom experiment to study the effect of material viscoelasticity on circumferential guided wave signals

Figure S2a shows the experiment set-up of the phantom experiment. The velocity map was extracted along the longitudinal path on the anterior wall. Figure S2b shows the experimental dispersion and its fitting result. The experimental dispersion curve was obtained by applying 2D-FFT to the velocity map. The axial guided wave model of a linear elastic tube immersed in fluid (*72*) was adopted to fit the experimental dispersion. Similar to the analysis of the *in vivo* experiment, the dominant wave mode excited by ARFs was assumed to be L(2,1). As a result, the Young's modulus of the phantom artery was fitted as 70.4 kPa by the L(2,1) mode. Figure S2b also draws L(1,1) mode by applying the aforementioned modulus. As shown, the dispersion of the L(1,1) mode is lower than that of the L(2,1) mode. We next adopted the FEA to verify the phantom experiment. The geometry and elastic modulus of the FEM was identical to the phantom. As depicted in Fig. S2c, the velocity map extracted along the longitudinal path was similar to that of the phantom experiment. The velocity map extracted along the circumferential path provides additional clarity, showing that the initial positive peak at the original position (indicated by the black arrow) represents the wavefront of the circumferential guided wave. Since the wave dispersion is more significant than that of the *in vivo* experiment, this wavefront cannot be directly used to calculate the circumferential wave velocity.

We then performed a tensile test to characterize the mechanical properties of the phantom. Uniaxial tension tests were conducted along both the axial and circumferential directions of the tube samples, respectively. As shown in Fig. S2d, the Young's modulus values were determined to be 65.5 kPa along the axial direction, and 66.4 kPa along the circumferential direction, respectively. The results of the tensile test indicate that the phantom is isotropic. By comparing the modulus obtained by the guided wave experiment and the tensile test, we show that the phantom possesses relatively good elasticity.

**Supplementary Note 3. Estimation of acoustic radiation force at the anterior and posterior walls**

The acoustic radiation force $f$ is proportional to the acoustic intensity $I$, and the acoustic intensity is proportional to the square of the acoustic pressure $P$ (*73*). Therefore, the ratio of the acoustic force between the anterior and posterior walls can be expressed as follows:

$$\frac{f_1}{f_2} = \left(\frac{P_1}{P_2}\right)^2 \tag{S1}$$

where $f_1$ and $f_2$ denote the acoustic radiation force at the anterior and posterior wall, respectively. $P_1$ and $P_2$ denote the acoustic pressure at the anterior and posterior wall, respectively. The acoustic pressure attenuates exponentially with depth in soft tissues, i.e.

$$P(z) = P_0 10^{-\alpha f_c z/20} \tag{S2}$$

where $\alpha$ denotes the acoustic attenuation. $f_c$ (= 7 MHz) denotes center frequency of the ultrasound transducer. $z$ denotes the distance from the ultrasound probe to the measured point. $P_0$ denotes the initial acoustic pressure emitted from the ultrasound probe. Inserting Eq. (S2) into Eq. (S1) yields

$$\frac{f_1}{f_2} = 10^{\alpha f_c \Delta z/10} \tag{S3}$$

where $\Delta z$ denotes the diameter of the artery.

For healthy volunteers, the acoustic attenuation of arteries is approximately 0.3 dB/cm/MHz (*73*). $\Delta z$ is approximately 0.7 cm. Using Eq. (S3) we get $f_1/f_2 \approx 1.4$. For hypertensive volunteers, the acoustic attenuation of arteries may even exceed 1 dB/cm/MHz (*74*). $\Delta z$ is approximately 0.9 cm. Using Eq. (S3) yields $f_1/f_2 \approx 4$.

**Supplementary Note 4. Modal analysis of the axial guided wave generated by the programmed ARFs**

The dispersion curve induced by the acoustic radiation force is supposed to be multimodal (*22*). In order to clarify the dominant modes in the axial guided wave generated by the proposed acoustic radiation force in this study, finite element analysis was conducted. To facilitate a comparison with theoretical models and without loss of generality, we built a tube model in vacuum with isotropic linear elastic material. The tube was firstly stimulated sequentially from the anterior wall to the posterior wall with Gaussian distribution of body force to simulate ARFs (Fig. S10a), and then stimulated at the anterior wall to simulate a single ARF (Fig. S10d). Their corresponding particle velocity maps extracted along the axial direction of the anterior wall are shown in Fig. S10b and e, respectively.

The theoretical axisymmetric L(0,1) and non-axisymmetric (flexural) modes L(N,1) (where $N \geq 1$, N denotes the number of periodic waves in the circumferential direction) of the isotropic tube in vacuum were calculated according to the literature (*75*). As for the excitation of ARFs, as shown in Fig. S10c, the dispersion curve is composed of multiple modes at lower frequencies (e.g. < 0.6 kHz), while at high frequencies (e.g. > 1 kHz), L(2,1) mode is the most dominant component and matches best to the dispersion curve. As for the excitation of a single ARF, as shown in Fig. S10f, both L(0,1) and L(1,1) modes match well to the dispersion curve at high frequencies. In summary, various excitation methods result in different dominant modes of guided waves. With the excitation method of ARFs, the L(2,1) mode predominates at high frequencies (e.g. > 1 kHz). Therefore, it is physically reasonable to use $N = 2$ in Eq. (2) to approximate the dispersion curve of axial guided waves.

It should be noticed that previous study (*69*) showed that an extended ARF beam (e.g. F-number = 3) selects mainly odd modes of axial guided waves (i.e. N = 1, 3, …). In our work, the acoustic radiation force is applied sequentially from the anterior to the posterior wall. Although a narrow Mach cone is formed, the F-number of each individual ARF is around 1.5, which explains why the dominant mode observed in our experiments is $N = 2$.

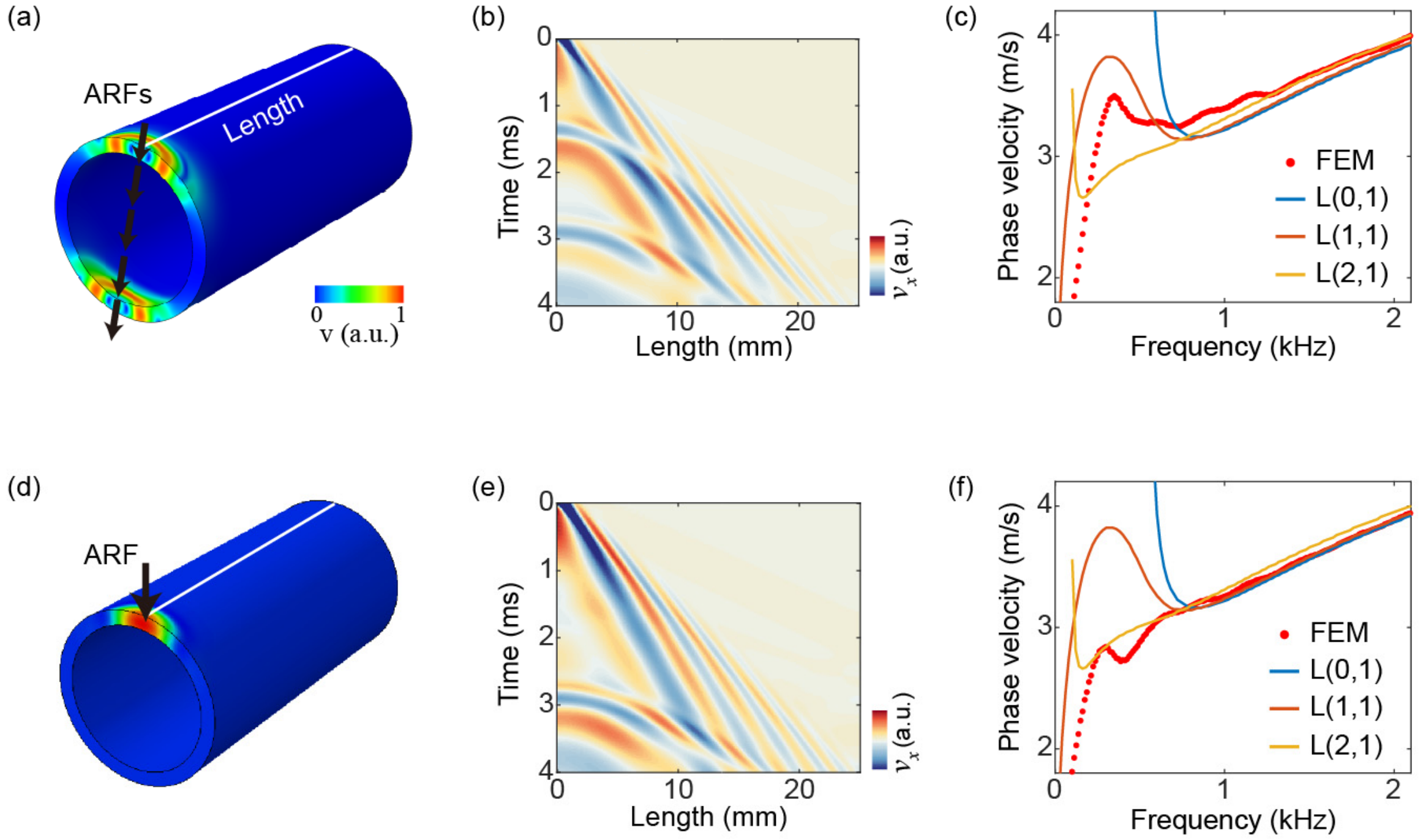

**Fig. S10**. Modal analysis of axial guided waves with different excitation methods. Excitation methods of (a) multiple ARFs or (d) a single ARF. Particle velocity maps extracted along the axial path in the anterior wall, including (b) multiple ARFs excitation, and (e) a single ARF excitation. Comparison of the simulated dispersion curve and first three theoretical modes of the axial guided waves, i.e. L(N,1) (N = 0,1,2), including (c) ARFs excitation, and (f) a single ARF excitation.

**Supplementary Note 5. Verification of the theoretical bidirectional guided wave model by FEA**

In this section, we compare the dispersion curves derived from the FEA model and those from the theoretical guided wave model. As shown in Fig. S11a, the dispersion curve predicted by the axial guided wave model with N = 2 aligns well with the that of the FEA in a frequency range of 0.6 – 1.3 kHz. On the contrary, applying N = 0 into Eq. (2) results in a dispersion curve that significantly deviates from the FEA results. This further demonstrates the necessity of accounting for the circumferential wavenumber in the approximate model of axial guided waves. Figure S11b compares the circumferential dispersion curves derived from the FEA and from the theoretical circumferential guided wave model. As shown, the theoretical curve is generally consistent with the FEA result.

In summary, the current axial guided wave model (Eq. (2) with N = 2) is appropriate to predict the axial dispersion of arteries in a frequency range of 0.6 – 1.3 kHz. The circumferential guided wave model (Eq. (3)) is appropriate to predict the circumferential dispersion of arteries over a wide frequency range (e.g. 0.3 – 1.5 kHz).

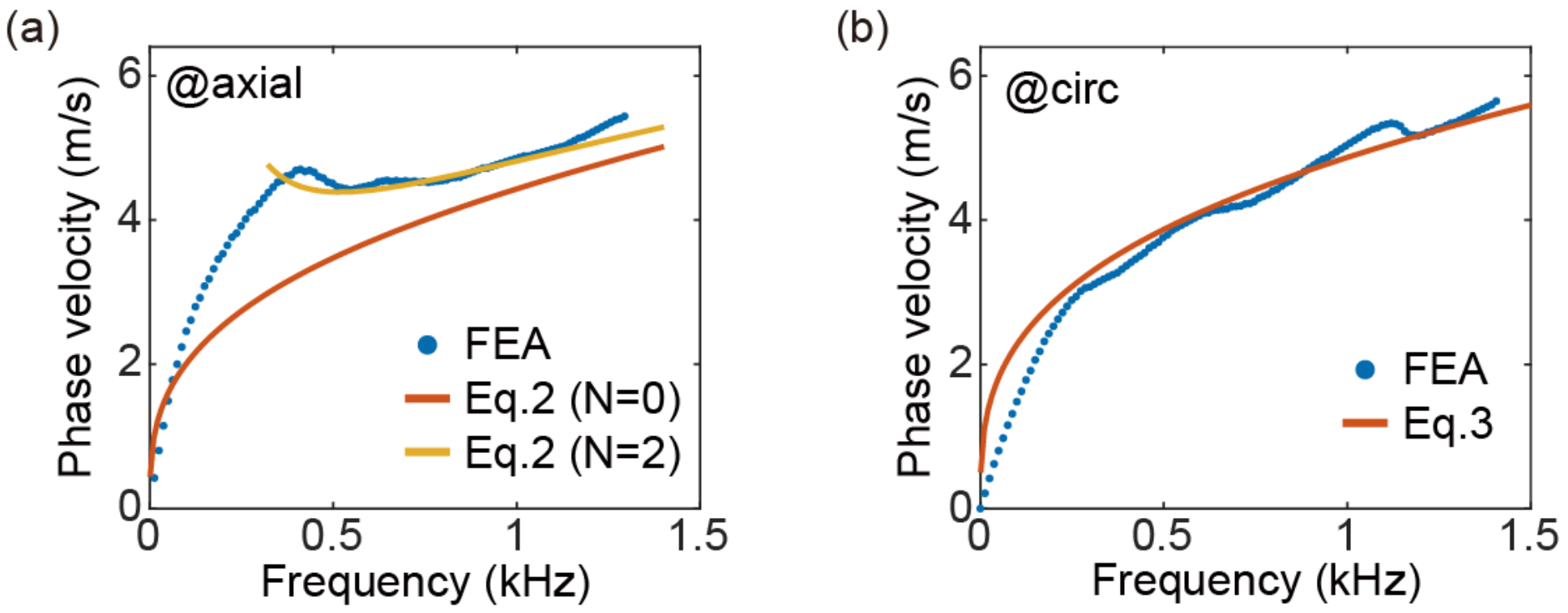

**Fig. S11**. Verification of the theoretical model by FEA. (a) Dispersion curve of axial guided waves. (b) Dispersion curve of circumferential guided waves.

**Supplementary Note 6. Mechanical meaning of the incremental parameter $\alpha$**

Consider the presence of a small shear deformation $k$ in a material subjected to the normal deformation, the deformation gradient tensor is

$$\boldsymbol{F} = \begin{bmatrix} \lambda_1 & k & \\ & \lambda_2 & \\ & & \lambda_3 \end{bmatrix} \tag{S4}$$

For the HGO model (*64*), the Cauchy stress of the shear component $\sigma_{12}$ is derived as

$$\sigma_{12} = 2W_1\lambda_2 k \tag{S5}$$

The tangent shear modulus $\mu_{12}^{'}$ can be defined as

$$\mu_{12}^{'} = \frac{\partial \sigma_{12}}{\partial k} \tag{S6}$$

Inserting Eq. (S5) into Eq. (S6), we obtain

$$\mu_{12}^{'} = 2W_1\lambda_2 \tag{S7}$$

Comparing Eq. (S7) and Eq. (S28-e), we find

$$\gamma = \mu_{12}^{'}\lambda_2 \tag{S8}$$

Inserting Eq. (S8) into relation $\sigma_1 = \alpha_1 - \gamma$, we get

$$\alpha_1 = \sigma_1 + \mu_{12}^{'}\lambda_2 \tag{S9}$$

Equation (S9) establishes a quantitative relationship between the incremental parameter, normal stress, and the tangent shear modulus. The incremental parameter $\alpha$ reflects a coupling effect of stress and material stiffness.

**Supplementary Note 7. The effect of windowing on the spatiotemporal velocity map to extract axial dispersion curves**

The spatiotemporal velocity field extracted along the longitudinal path contains both axial and circumferential guided wave information. We adopted the FEA results to study the effect of windowing on the map to extract axial guided wave dispersion. Figure S12a shows the raw particle velocity map. Figure S12b shows the frequency-wavenumber field ($k$-space) by applying 2D-FFT to the raw spatiotemporal map, where the spectral signal of the circumferential guided wave is very distinct. Figure S12c and d demonstrate that applying windowing to the velocity map, followed by a 2D-FFT on the map, results in a k-space map primarily representing the axial guided wave components, with the circumferential wave components effectively suppressed. The dispersion curve is then obtained by searching peaks in the $k$-space map at each frequency, resulting in a frequency-wavenumber curve (red dotted line in Fig. S12b and d), and the phase velocity is calculated by $c_a^p = f/k$. Figure S12e compares the two dispersion curves, obtained with or without the use of windowing. The one with windowing is smoother, highlighting the necessity of windowing. Figure S12f - h show the extraction of axial guided wave dispersion from the *in vivo* data. As shown, by windowing the spatiotemporal map, the circumferential guided wave signals can be effectively filtered out, and the resulting axial wave dispersion curve is smoother.

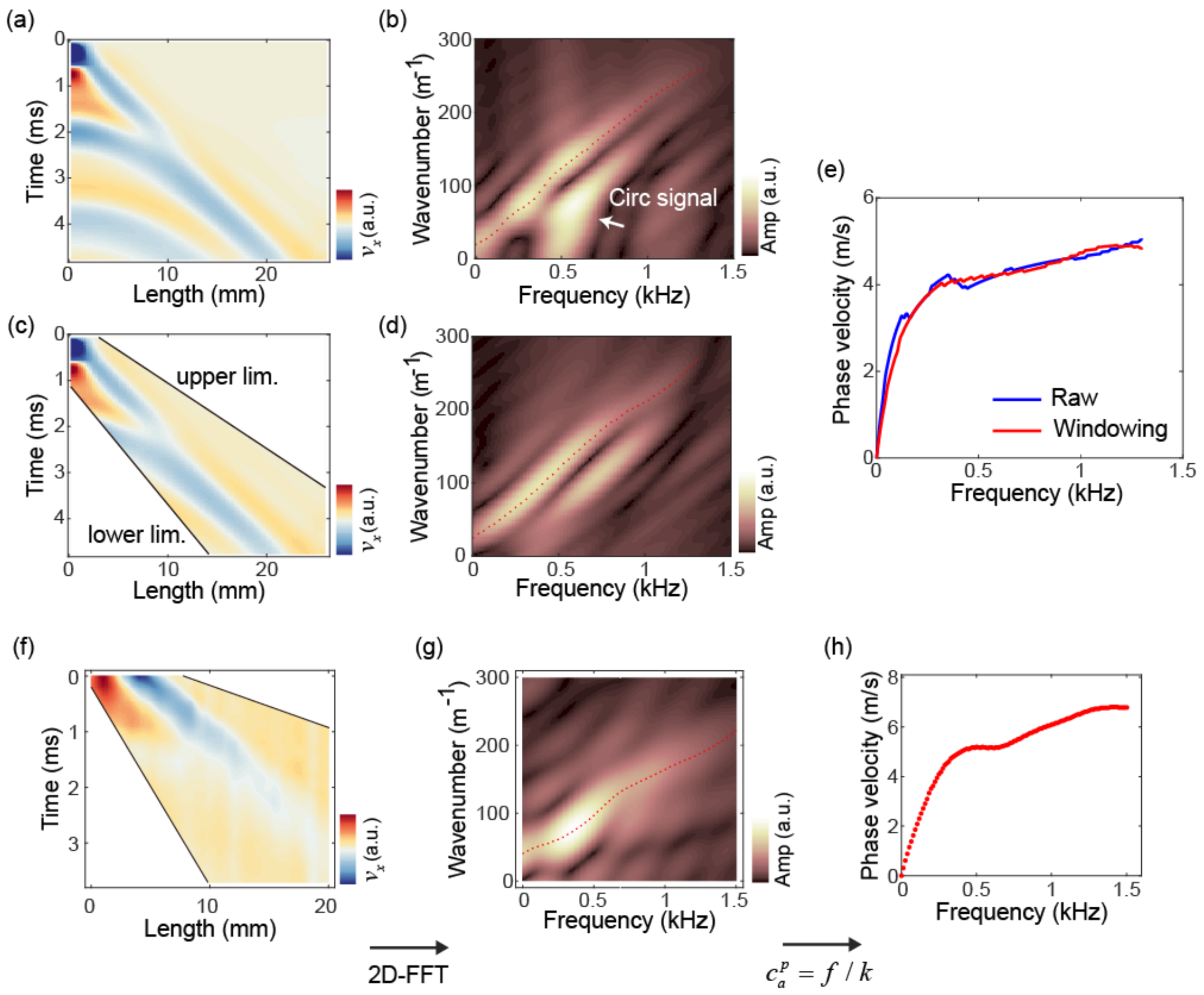

**Fig. S12**. Extraction of axial guided wave dispersion curves. (a)-(e): finite element results. (f)-(h): *in vivo* data. (a) Raw spatiotemporal velocity map, and (b) the corresponding *k*-space by using 2D-FFT. (c) Windowing on the spatiotemporal velocity map to filter circumferential signal, and (d) the corresponding *k*-space by using 2D-FFT. (e) Dispersion curves obtained from the raw velocity map and the windowing map, respectively. (f) Windowing on the *in vivo* velocity map, (g) the corresponding *k*-space map, and (h) the dispersion curve.

**Supplementary Note 8. *Ex vivo* experiment to verify the feasibility of the inversion method for characterizing arterial mechanical stress and stiffness**

**8.1 Guided wave elastography for axial stress characterization using the inversion method**

Figure S13a shows the experiment set-up of the guided wave elastography on the porcine aorta. Figure S13b plots the velocity maps extracted from the anterior wall at the axial stretch ratio of 1 (stress-free), 1.23 and 1.31, respectively. Using the axial dispersion data at $\lambda$ = 1.23 and 1.31 as input for the inversion method, we identified the parameters $\alpha_{a,\mathrm{d}}$, $\alpha_{a,\mathrm{s}}$, $\gamma_\mathrm{d}$, $\gamma_\mathrm{s}$, $g$ and $\tau$, where the subscripts 'd' and 's' refer to the state of $\lambda$ = 1.23 and 1.31, respectively. These values are listed in Table. S2. Figure S13c shows the dispersion data at the two states and their fitting curves. The axial stresses at the two states were then calculated and compared with the results from the tensile tests. As shown in Fig. S13d, the axial stress can be obtained stably by the inversion, with a relative error of less than 9%.

Figure S14a plot two stress - stretch curves, obtained from uniaxial tensile test along the axial and circumferential directions of the aorta samples, respectively. By fitting to the HGO model (*64*), the constitutive parameters of the aorta can be obtained: $\mu = 38.3$ kPa , $k_1 = 111.2$ kPa , $k_2 = 11.5$ , $\kappa = 0.202$ , $\varphi = 39.47°$ . Figure S14b shows the dispersion curve at the stress-free state. The two viscous parameters $g$ and $\tau$ can be fitted to the theoretical dispersion: $g$ = 0.66 and $\tau = 0.077$ms.

Taking the above constitutive parameters into Eq. (S28), we were able to predict the incremental parameters $\alpha_a$ and $\gamma$, and compare them to the values obtained by the inversion method. According to Table S2, the parameters $\alpha_a$, $\gamma$, and $g$ can be inversed in a relatively stable way, with relative error less than 10%. On the other hand, the relative error of $\tau$ is high, indicating that the current inversion method is not reliable enough to characterize arterial viscosity.

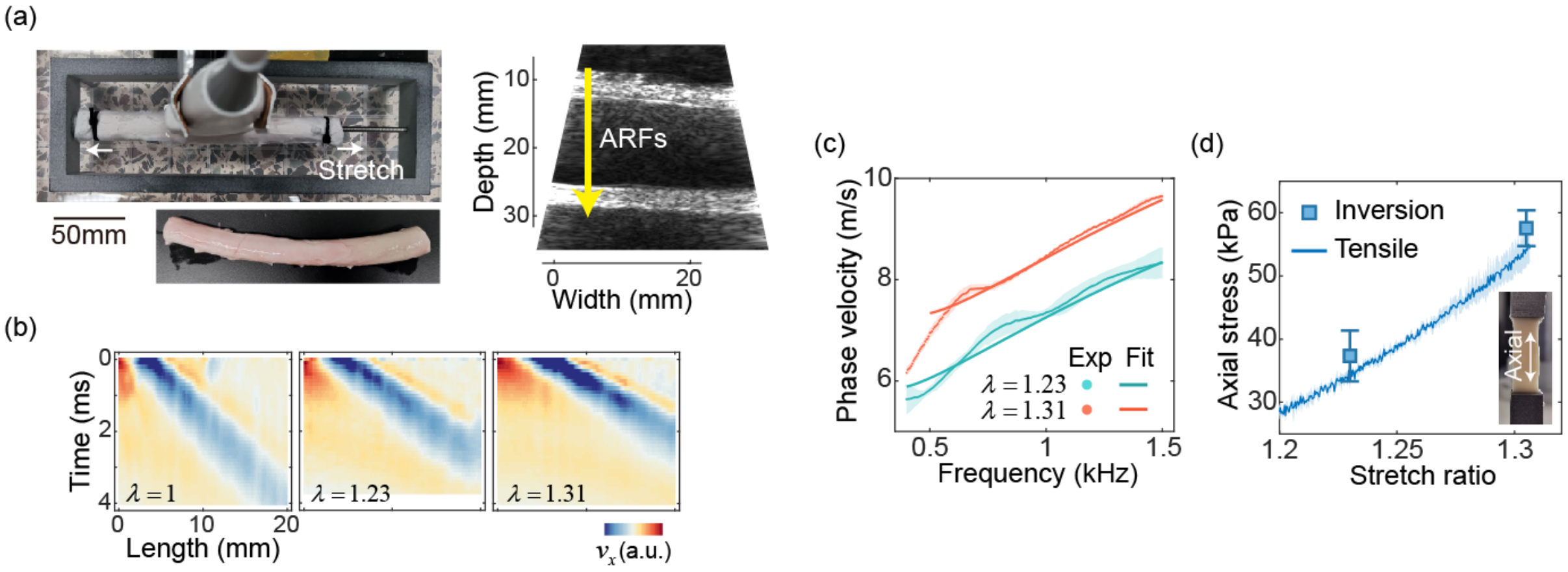

**Fig. S13**. Results of the guided wave elastography on the porcine aorta. (a) Experimental set-up of the elastography. (b) Velocity maps at axial stretch of $\lambda$ = 1, 1.23 and 1.31. (c) Dispersion curves and fitting results at the axial stretch ratio of 1.23 and 1.31, respectively. (d) Comparison of the axial stresses obtained by the inversion method and the tensile test.

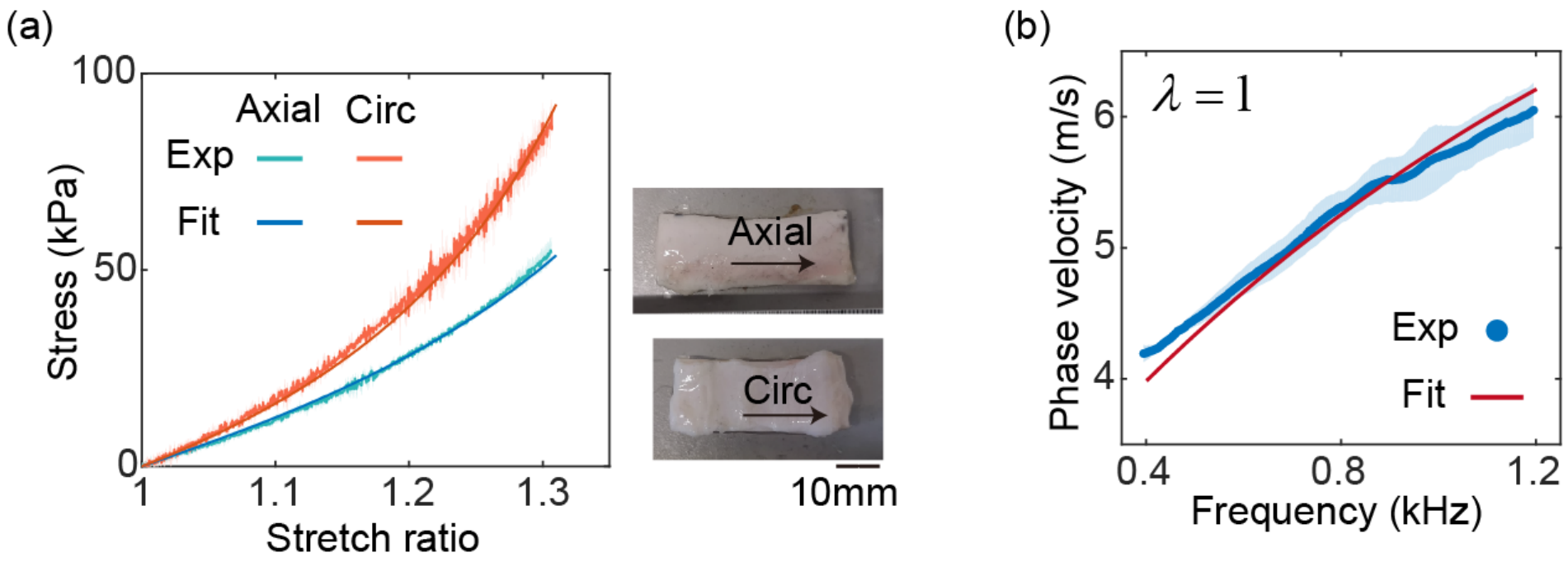

**Fig. S14**. Mechanical characterization of the porcine aorta. (a) Tensile tests along the axial and circumferential of the aorta. (b) Axial dispersion at the stress-free state and its fitting curve.

**8.2 Discussion on the characteristics of the time phase $t_1$ in the velocity map**

It should be noted that from the velocity map of the *ex vivo* aorta, the marker to measure circumferential wave velocity – time phase $t_1$ is less obvious to be detected; this is different from what we have observed in the *in vivo* experiments. We assume that this can be attributed to the fact that the radius of the porcine aorta is much larger than the that of the human carotid artery. As a result, the travel time of the circumferential guided wave from the posterior wall to the anterior wall may be longer, and the attenuation of this wave could be more pronounced, leading to a less distinct arrival signal. Figure S15a shows the velocity map of the *ex vivo* experiment at the stress-free state. The time phase $t_1$ can be detected at ~ 4 ms, where the velocity peak is relatively weak. Given that the imaging time of the *ex vivo* experiment is no more than 4 ms, it is not surprising that the time $t_1$ cannot be observed in the map.

We also verify this assumption by FEA. Through the tensile test, we were able to characterize the anisotropy of the aorta within the linear deformation range. As shown in Fig. S15c, the axial and circumferential Young's modulus of the aorta are $E_a = 116.1\,\text{kPa}$ and $E_c = 197.7\,\text{kPa}$, respectively. Taking the material parameters and aortic geometry into FEA model, we obtained the velocity map calculated by the FEA (Fig. S15b). The time phase $t_1$ is basically consistent with the experimental result, thus confirming the assumption.

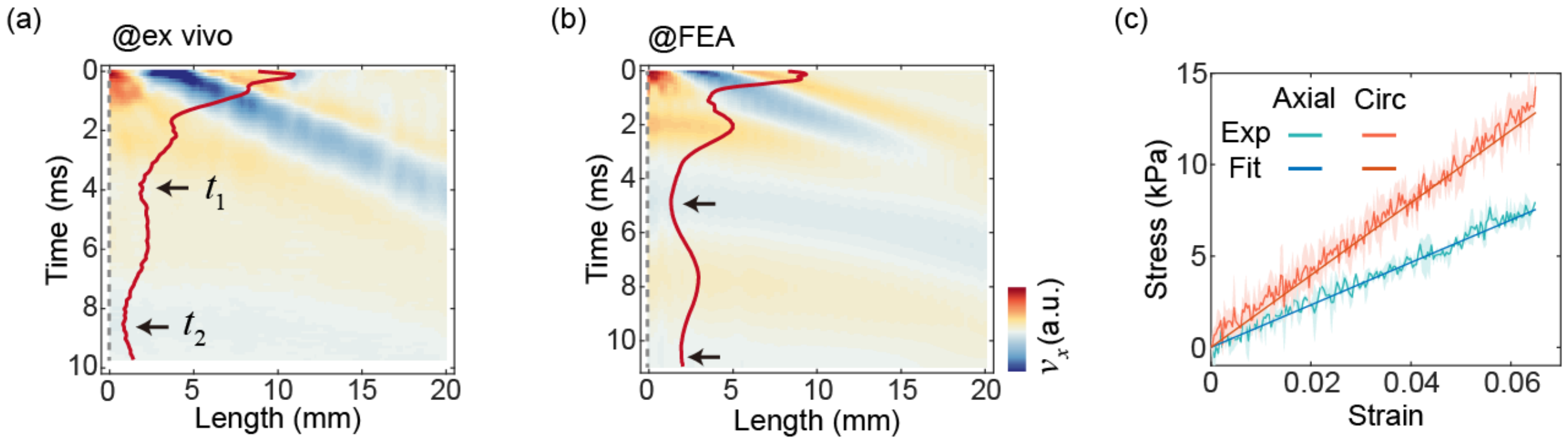

**Fig. S15**. (a) Particle velocity map of the *ex vivo* experiment at the stress-free state. (b) Particle velocity map of the FEA result. Material parameters include $E_a = 116.1\ \text{kPa}$, $E_c = 197.7\ \text{kPa}$, $\mu_a = 38.3\ \text{kPa}$ and $\mu_c = 87\ \text{kPa}$. (c) Tensile test to characterize the material anisotropy of the aorta within the linear deformation range.

**Supplementary Note 9. The impact of compensation time selection on the measurement error of circumferential wave velocities**

In this section, we discuss the measurement error of circumferential wave velocities. According to Eq. (4), an appropriate selection of the compensation time $t_c$ is crucial for the measurement of circumferential wave velocities. In our *in vivo* experiments, the ARFs were sequentially induced from the anterior wall to the posterior wall, resulting in a time delay between the guided waves excited on the anterior and posterior walls. Therefore, it is necessary to examine the impact of this time delay, which optimizes the selection of time $t_c$. In the following, we firstly determine the spatial distribution of a single ARF. Then we conduct FEA with the known spatial distribution and temporal sequence of ARFs, aiming to quantify the time delay between two walls and study the measurement error of circumferential wave velocities.

As shown in Fig. S16a, the spatial distribution of ARF can be expressed in form of Gauss function $f = f_0 \exp\left(-x^2 / 2r_x^2 - y^2 / 2r_z^2\right)$, where $r_x$ and $r_z$ represent the Gauss radius in the x (the axial direction of ultrasound beam) and z axis (the lateral direction of ultrasound beam), respectively. The origin corresponds to the acoustic focus location. We have experimentally calibrated the Gauss radius along the *z* axis in our previous work (*20*), i.e. $r_z = 0.2$ mm (Fig. S16b). In order to determine the Gauss radius along the x axis, we built a 2D FE model with ~2,000,000 acoustic elements. Similar to the ultrasound settings in the experiment, 32 elements (spanning ~10 mm) were sinusoidally driven with specific time delays for each element to ensure that all acoustic waves would converge at the depth of 15 mm. The simulation of the acoustic field provides a ratio between $r_x$ and $r_z$ of ~10 (Fig. S16a). Finally, we derived that the Gauss radius $r_x$ was approximately 2 mm.

By configuring the spatial distribution of ARF with the specified value, and sequentially exciting 10 ARFs within 0.5 ms, we simulated the propagation of guided waves in arteries by FEA (Fig. S16c). Figure S16d displays the spatiotemporal velocity maps extracted from the anterior wall, the posterior wall, and the circumferential path within half a circle. The time phases $t_1$ detected at the anterior wall ($t_{1,a}$) and at the posterior wall ($t_{1,p}$) show a time lag of ~0.2 ms. It is reasonable because the anterior wall is excited in advance compared to the posterior wall by ARFs, resulting in the circumferential guided wave arriving at the posterior wall earlier than it does at the anterior wall. By measuring the slope (black dash line) in the map extracted along the circumferential path, we obtained the truth value of the circumferential wave velocity $c_c$. In principle, the velocity $c_c$ can be calculated using either of the following two equations,

$$c_c = \pi r_c / \left( t_{1,a} + t_{c,a} \right) \quad \text{(S10)}$$

$$c_c = \pi r_c / \left( t_{1,p} + t_{c,p} \right) \quad \text{(S11)}$$

where $t_{c,a}$ and $t_{c,p}$ denote the compensation time for use with the anterior wall and posterior wall, respectively. In our *in vivo* experiments, we adopted $t_c = 0.8$ ms for both the anterior wall and posterior walls, which brings the measurement error of $c_c$. As illustrated in Fig. S16e, when both $t_{c,a}$ and $t_{c,p}$ are set to 0.8 ms, compared to the truth value of $c_c$, the calculation value of $c_c$ from the anterior wall by Eq. (S10) will slightly underestimate it, whereas the calculation from the posterior wall by Eq. (S11) will slightly overestimate it. However, the relative error for both of them is less than 6%, indicating that the measurement error of circumferential wave velocities using the current method is negligible.

From the viewpoint of the experimental measurement, the temporal resolution of ultrasound elastography is 0.1 ms, therefore it's challenging to accurately measure the time difference between $t_{1,a}$ and $t_{1,p}$. Figure S16f displays *in vivo* results from one representative volunteer, from which both the $t_{1,a}$ and $t_{1,p}$ were measured in the corresponding maps. By statistical analysis, no significant difference was observed between the two time phases in cardiac cycles. This result verifies that the temporal resolution of the experiment is insufficient to distinguish the time difference between $t_{1,a}$ and $t_{1,p}$ accurately. This also illustrates the rationale behind the experimental choice of uniformly setting $t_c$ to 0.8 ms for both the anterior and posterior walls.

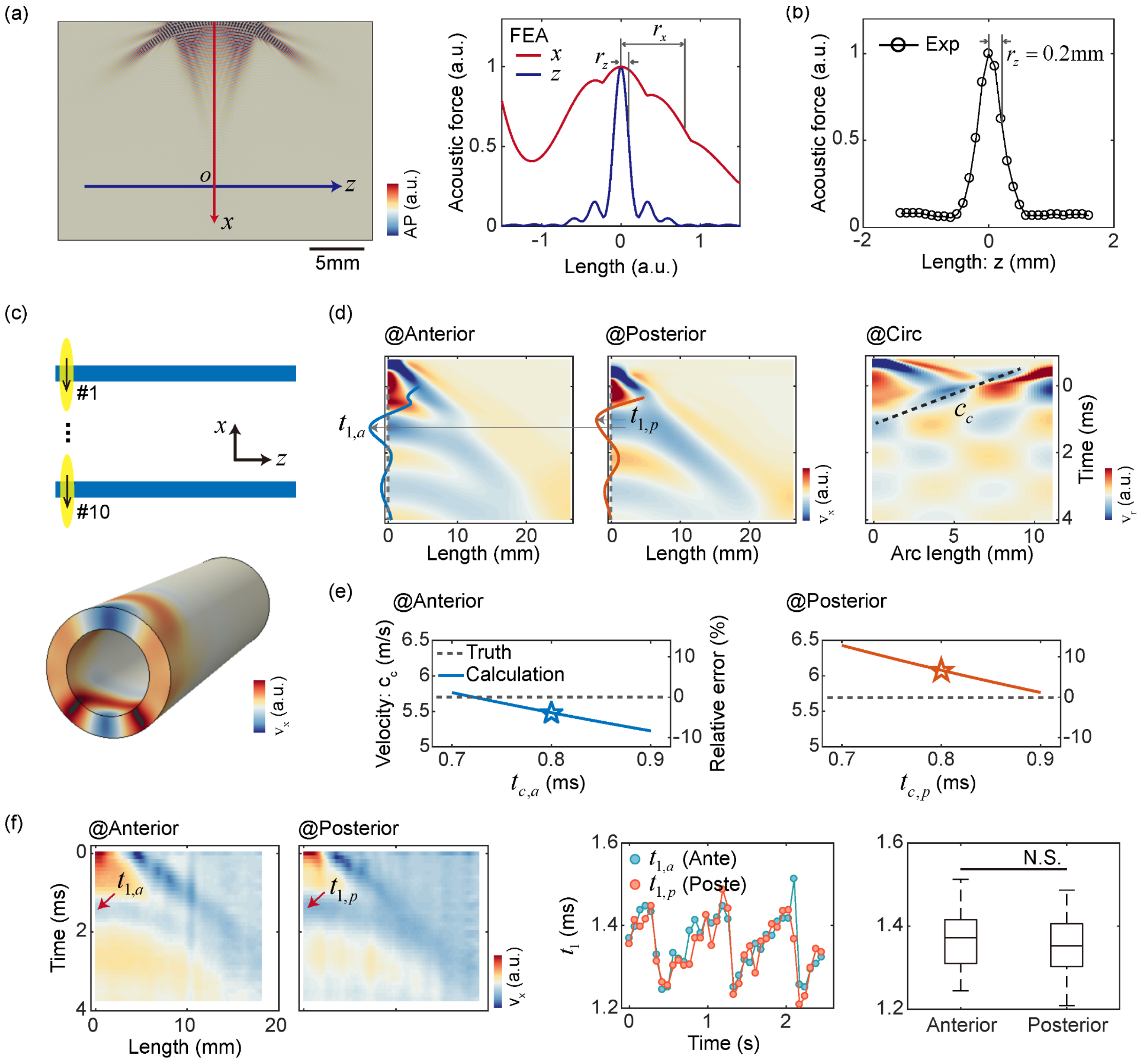

**Fig. S16**. A discussion on the measurement error of circumferential wave velocities. (a) FEA to simulate the acoustic field distribution by ARF. (b) Experimental measurement of the acoustic field along the z axis. (c) FEA to simulate arterial guided waves. (d) Particle velocity maps extracted from the anterior wall, the posterior wall and the circumferential path. (e) Relative error of the measurement of $c_c$ from the anterior wall and posterior wall. (f) *In vivo* experimental results from a representative volunteer, including particle velocity maps at two walls, variation of $t_1$ in cardiac cycles, and comparison of $t_{1,a}$ and $t_{1,p}$.

**Supplementary Note 10. Comparison of the two viscoelastic models**

Biomaterials exhibit a power-law viscoelastic response. Compared with the Prony series model adopted in this study, it is more appropriate to describe the viscoelasticity of arterial walls by the Kelvin-Voigt fractional derivative model (KVFD model) (*66*). The constitutive parameters of the KVFD model include fractional order $\beta_0$ ($0 < \beta_0 < 1$), and viscous parameter $\eta$ (dimension [$s^{\beta_0}$]). In order to compare the two viscoelastic models and illustrate the feasibility of using the Prony series model in this study, we compare the dispersion curves of both models (*71*). As shown in Fig. S17, when the frequency is below 1.5 kHz, the two dispersion curves are basically consistent. This suggests that the first-order Prony series model can effectively capture the viscoelastic behavior of the arteries within the frequency range measured by ultrasound elastography. As the frequency increases, for example, exceeding 10 kHz, the dispersion of the KVFD model continues to rise steadily, while the dispersion of the Prony series model reaches plateaus. Therefore, the first-order Prony series model employed in this study can sufficiently characterize the dispersion within the frequency range measured by ultrasound elastography. However, the corresponding viscoelastic parameters may not be applicable for predicting the dispersion of arteries at high frequencies (e.g. over 5 kHz as detected by optical coherence elastography).

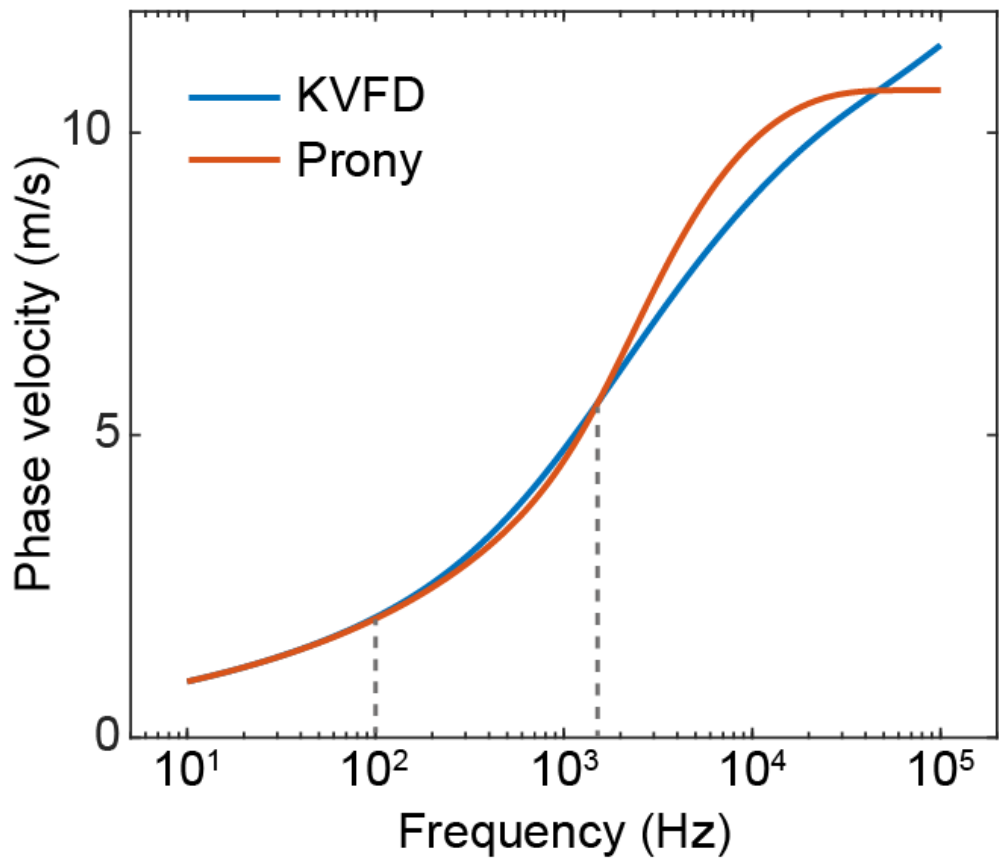

**Fig. S17**. Comparison of the two viscoelastic models. The parameters of the KVFD model are $\eta = 0.1\,\mathrm{s}^{\beta_0}$, $\beta_0 = 0.25$. The parameters of the Prony series model are $g = 0.7$, $\tau = 6\times10^{-5}$ s.

**Supplementary Note 11. Guided wave model for pre-stressed viscoelastic flat plates**

Consider a viscoelastic pre-stressed flat plate immersed in inviscid fluid (Fig. S9b). A Cartesian coordinate system is defined on the plate, with the three directions referred to as axial ($x_z$), circumferential ($x_\theta$), and radial directions ($x_r$). The wall thickness is $h$. In the incremental dynamics theory of viscoelastic materials, the incremental stress $\mathbf{\Sigma}$ is related to the displacement of wave motion $\boldsymbol{u}$ as follows (*71*)

$$\Sigma_{ji} = -\hat{p}\delta_{ji} + pu_{j,i} - G\hat{Q}\delta_{ji} + GQu_{j,i} + G\mathcal{A}_{0jikl}u_{l,k} - \Omega\sigma^e_{Djk}u_{i,k} \tag{S12}$$

where $j$ and $i$ correspond to the directional indices $r$, $\theta$ and $z$. $\hat{p}$ denotes the increment of the Lagrange multiplier $p$. $\hat{Q}$ is the increment of the volumetric part of the elastic stress $Q$ $(= \sigma^e_{ii}/3)$. $\mathcal{A}_{0jikl} = \left(\partial^2 W / \partial F_{i\alpha}\partial F_{l\beta}\right)F_{j\alpha}F_{k\beta}$ is the fourth-order Eulerian elasticity tensor. $G$ and $\Omega$ are two frequency-dependent parameters

$$G = \left(1 - \frac{g}{1+i\omega\tau}\right)/\left(1-g\right),\ \Omega = \left(\frac{g\omega^2\tau^2}{1+\omega^2\tau^2} + i\frac{g\omega\tau}{1+\omega^2\tau^2}\right)/\left(1-g\right) \tag{S13}$$

where $i$ denotes the imaginary unit. Viscous parameters $g$ and $\tau$ denote the relaxation scale and characteristic relaxation time of arteries, respectively. $\omega$ $(= 2\pi f)$ denotes the angular frequency.

When the guided wave propagates along the axial direction, assuming that the wave has displacement components only within the $x_z$-$x_r$ plane (i.e. $u_\theta = 0$), a stream function $\psi(x_z, x_r, t)$ can be used to replace displacements $u_z = \psi_{,r}$ and $u_r = -\psi_{,z}$. Inserting Eq. (S12) and $\psi$ into Eq. (1), the wave motion equation can be simplified as

$$G\left(\alpha_a\psi_{,zzzz} + 2\beta_a\psi_{,zzrr} + \gamma\psi_{,rrrr}\right) - \Omega\left[\sigma^e_{Dzz}\psi_{,zzzz} + \sigma^e_{Drr}\psi_{,rrrr} + (\sigma^e_{Dzz} + \sigma^e_{Drr})\psi_{,zzrr}\right] = \rho(\psi_{,zztt} + \psi_{,rrtt}) \tag{S14}$$

where

$$\sigma^e_{Dzz} = \tfrac{2}{3}\alpha_a - \tfrac{1}{3}\gamma - \tfrac{1}{3}\alpha_c,\ \sigma^e_{Drr} = \tfrac{2}{3}\gamma - \tfrac{1}{3}\alpha_a - \tfrac{1}{3}\alpha_c \tag{S15}$$

The long-term incremental parameters $\alpha_a$, $\alpha_c$, $\gamma$, $\beta_a$ are defined as $\alpha_a = \mathcal{A}_{0zrzr}$, $\alpha_c = \mathcal{A}_{0\theta r\theta r}$, $\gamma = \mathcal{A}_{0rzrz} = \mathcal{A}_{0r\theta r\theta}$ , $\beta_a = (\mathcal{A}_{0zzzz} + \mathcal{A}_{0rrrr} - 2\mathcal{A}_{0zzrr} - 2\mathcal{A}_{0zrrz})/2$ . Since the elastic wave propagates along the $x_z$ axis, the stream function has a harmonic form of $\psi = \psi_0\exp(skx_r)\exp[i(kx_z - \omega t)]$, where $\psi_0$ is an amplitude. $s$ denotes a ratio of wavenumbers in the two directions. $k$ denotes the complex wavenumber. Inserting $\psi$ into Eq. (14) yields

$$(\gamma+\Omega Q)s^4+\left[\rho\frac{\omega^2}{k^2}-2G\beta_a+\Omega(\alpha_a+\gamma-2Q)\right]s^2+\alpha_a+\Omega Q-\rho\frac{\omega^2}{k^2}=0 \tag{S16}$$

where

$$Q=(\alpha_a+\gamma+\alpha_c)/3 \tag{S17}$$

At the interface between the plate and fluid, the conditions of normal displacement continuity, normal stress continuity, and zero shear stress are satisfied, which can be expressed as:

$$u_r=u_r^f,\Sigma_{rz}=0,\Sigma_{rr}=-p^f,\text{at } x_r=\pm h/2 \tag{S18}$$

where $u^f$ denotes the displacement of fluid. $p^f$ denotes the hydrostatic pressure of fluid. Combining Eq. (12) and boundary conditions Eq. (18), we can derive the dispersion equation for the antisymmetric mode of guided waves as follows (*71*)

$$\begin{aligned}&\left(1+s_{2a}{}^2\right)\cdot\left(-\rho\frac{\omega^2}{k^2}s_{1a}+C_{1a}s_{1a}-C_{2a}s_{1a}{}^3\right)\cdot\tanh\left(s_{1a}kh/2\right)\\&-\left(1+s_{1a}{}^2\right)\cdot\left(-\rho\frac{\omega^2}{k^2}s_{2a}+C_{1a}s_{2a}-C_{2a}s_{2a}{}^3\right)\cdot\tanh\left(s_{2a}kh/2\right)+\left(s_{1a}{}^2-s_{2a}{}^2\right)\frac{\rho^f}{\xi}\frac{\omega^2}{k^2}=0\end{aligned} \tag{S19}$$

where $s_{1a}$ and $s_{2a}$ are two roots solved by the quartic equation (S16), and

$$C_{1a}=2G\beta_a+\gamma+\Omega\alpha_c \tag{S20}$$

$$C_{2a}=\gamma+\Omega Q \tag{S21}$$

$$\xi^2=1-\frac{\omega^2}{k^2}\frac{1}{c_f{}^2} \tag{S22}$$

where $\rho$ and $\rho^f$ denote the density of the arterial wall (1000 kg/m$^3$) and blood (1000 kg/m$^3$), respectively. The speed of sound in the fluid is $c_f=\sqrt{\kappa_p/\rho^f}$ . $\kappa_p$ denotes the bulk modulus of the fluid (= 2.2 GPa for blood). Parameter sensitivity analysis indicates that the parameters $\alpha_c$ and $\beta_a$ have a minimal effect on axial guided waves (see details in Supplementary Note 15). Therefore, we can insert $\alpha_c=1.5\alpha_a$ and $\beta_a=4\alpha_a$ into Eq. (S19), and as a result we obtain Eq. (10) used in the main text. The rationale for using the relations $\alpha_c=1.5\alpha_a$ and $\beta_a=4\alpha_a$ is supported by experimental observations and numerical calculations (see Supplementary Note 13).

As for the waves propagating along circumferential direction, the dispersion equation for the antisymmetric mode of guided waves can be derived in a similar way as mentioned above, which gives

$$
\begin{aligned}
&\left(1+{s_{2c}}^2\right)\cdot\left(-\rho\frac{\omega^2}{k^2}s_{1c}+C_{1c}s_{1c}-C_{2c}{s_{1c}}^3\right)\cdot\tanh\left(s_{1c}kh/2\right)\\
&-\left(1+{s_{1c}}^2\right)\cdot\left(-\rho\frac{\omega^2}{k^2}s_{2c}+C_{1c}s_{2c}-C_{2c}{s_{2c}}^3\right)\cdot\tanh\left(s_{2c}kh/2\right)+\left({s_{1c}}^2-{s_{2c}}^2\right)\frac{\rho^f}{\xi}\frac{\omega^2}{k^2}=0
\end{aligned}
\tag{S23}
$$

where $s_{1c}$ and $s_{2c}$ are two roots solved by the quartic quation

$$
\left(\gamma+\Omega Q\right)s^4+\left[\rho\frac{\omega^2}{k^2}-2G\beta_c+\Omega\left(\alpha_c+\gamma-2Q\right)\right]s^2+\alpha_c+\Omega Q-\rho\frac{\omega^2}{k^2}=0 \tag{S24}
$$

and

$$
C_{1c}=2G\beta_c+\gamma+\Omega\alpha_a \tag{S25}
$$

$$
C_{2c}=\gamma+\Omega Q \tag{S26}
$$

where long-term incremental parameter $\beta_c$ is defined as $\beta_c=(\mathcal{A}_{0\theta\theta\theta\theta}+\mathcal{A}_{0rrrr}-2\mathcal{A}_{0\theta\theta rr}-2\mathcal{A}_{0\theta rr\theta})/2$. Given that the parameters $\alpha_a$ and $\beta_c$ have a minimal effect on circumferential guided waves, by inserting $\alpha_a=0.7\alpha_c$ and $\beta_c=4\alpha_c$ into Eq. (S23), we can obtain Eq. (15) in the main text. The rationale for using the relations $\alpha_a=0.7\alpha_c$ and $\beta_c=4\alpha_c$ is supported by experimental observations and numerical calculations (see Supplementary Note 13).

## Supplementary Note 12. Explicit forms of incremental parameters

The general definition of the fourth-order Eulerian elasticity tensor $\mathcal{A}_{0jikl}$ can be found in previous literature (*76*). In particular, for the HGO model (see Eq. (8) for the strain energy function *W*), the elasticity tensor has the following explicit form,

$$\begin{aligned}\mathcal{A}_{0jikl} &= 2W_1\delta_{il}B_{jk} + 2W_4\delta_{il}m_jm_k + 2W_6\delta_{il}m_j^{'}m_k^{'} \\ &+ 4W_{11}B_{ij}B_{kl} + 4W_{14}\left(B_{ij}m_km_l + B_{kl}m_im_j\right) \\ &+ 4W_{16}\left(B_{ij}m_k^{'}m_l^{'} + B_{kl}m_i^{'}m_j^{'}\right) + 4W_{44}m_im_jm_km_l \\ &+ 4W_{66}m_i^{'}m_j^{'}m_k^{'}m_l^{'} + 4W_{46}\left(m_im_jm_k^{'}m_l^{'} + m_i^{'}m_j^{'}m_km_l\right) \qquad \{i,j,k,l\} \in \{r,\theta,z\}\end{aligned}$$

(S27)

where $W_i = \partial W / \partial I_i$, $W_{ij} = \partial^2 W / \partial I_i \partial I_j$, $\boldsymbol{B} = \boldsymbol{F}\boldsymbol{F}^{\mathrm{T}}$, $\boldsymbol{m} = \boldsymbol{m}_1 = \boldsymbol{F}\boldsymbol{M}_1$, $\boldsymbol{m}' = \boldsymbol{m}_2 = \boldsymbol{F}\boldsymbol{M}_2$. $\boldsymbol{F}$ (= diag($\lambda_r, \lambda_\theta, \lambda_z$)) denotes the deformation gradient tensor. $\boldsymbol{M}_1$ and $\boldsymbol{M}_2$ are two symmetric fiber orientations defined in the initial state, each at an angle $\varphi$ relative to the circumferential direction ($\theta$-axis). The incremental parameters involved in the theoretical dispersion of guided waves have the following explicit forms

$$\alpha_a = 2W_1\lambda_z^2 + 2W_4\lambda_z^2\sin^2\varphi + 2W_6\lambda_z^2\sin^2\varphi \tag{S28-a}$$

$$\begin{aligned}\beta_a &= W_1\left(\lambda_z^2 + \lambda_r^2\right) + W_4\lambda_z^2\sin^2\varphi + W_6\lambda_z^2\sin^2\varphi + 2W_{11}\left(\lambda_z^2 - \lambda_r^2\right)^2 \\ &+ 4W_{14}\lambda_z^2\sin^2\varphi\left(\lambda_z^2 - \lambda_r^2\right) + 4W_{16}\lambda_z^2\sin^2\varphi\left(\lambda_z^2 - \lambda_r^2\right) \\ &+ 2W_{44}\lambda_z^4\sin^4\varphi + 2W_{66}\lambda_z^4\sin^4\varphi\end{aligned} \tag{S28-b}$$

$$\alpha_c = 2W_1\lambda_\theta^2 + 2W_4\lambda_\theta^2\cos^2\varphi + 2W_6\lambda_\theta^2\cos^2\varphi \tag{S28-c}$$

$$\begin{aligned}\beta_c &= W_1\left(\lambda_\theta^2 + \lambda_r^2\right) + W_4\lambda_\theta^2\cos^2\varphi + W_6\lambda_\theta^2\cos^2\varphi + 2W_{11}\left(\lambda_\theta^2 - \lambda_r^2\right)^2 \\ &+ 4W_{14}\lambda_\theta^2\cos^2\varphi\left(\lambda_\theta^2 - \lambda_r^2\right) + 4W_{16}\lambda_\theta^2\cos^2\varphi\left(\lambda_\theta^2 - \lambda_r^2\right) \\ &+ 2W_{44}\lambda_\theta^4\cos^4\varphi + 2W_{66}\lambda_\theta^4\cos^4\varphi\end{aligned} \tag{S28-d}$$

$$\gamma_a = \gamma_c = \gamma = 2W_1\lambda_r^2 \tag{S28-e}$$

where $\lambda_\theta$, $\lambda_z$ and $\lambda_r$ denote circumferential, axial, and radial stretch ratio, respectively.

**Supplementary Note 13. Parameter space for the inverse parameters**

The dispersion equation Eq. (S19) contains six unknown parameters, i.e., nonlinear elastic parameters $\alpha_a$, $\alpha_c$, $\beta_a$, $\gamma$, and viscous parameters $g$, $\tau$. In order to understand the relationship between the four elastic parameters and also determine the parameter space, we varied constitutive parameters of HGO model, as well as blood pressures over a sufficiently wide range, i.e. constitutive parameters: $5 < \mu < 50$ kPa , $10 < k_1 < 500$ kPa , $0 < k_2 < 100$ , $0 < \varphi < \pi / 2$ , $0 < \kappa < 1/3$ (*46*, *49, 50*); blood pressure: $40 \le P \le 250$ mmHg . In practice, we firstly adopted the thin-walled model to determine the circumferential stretch ratio $\lambda_\theta$ of arteries with respect to blood pressure $P$ (*77*),

$$P = \frac{H}{R_c} \frac{1}{\lambda_\theta \lambda_z} \frac{\partial W}{\partial \lambda_\theta} \tag{S29}$$

where $R_c$ and $H$ denote the radius and wall thickness of the artery at the unloaded state. Axial stretch ratio is confined in a range of $1 \le \lambda_z \le 1.2$ (*48-49*). Then by using Eqs. (S28) and (S29), we determined the parameter space of the incremental parameters. They are $20 \le \alpha_a \le 200$ kPa , $0 < \gamma / \alpha_a < 1$, $\beta_a / \alpha_a > 2$ and $\alpha_c / \alpha_a > 0.8$.

We further investigated the parameter space of the viscoelastic parameters $g$ and $\tau$. From FEA, we found material viscoelasticity played a crucial role in shaping the spatiotemporal map. As shown in Fig. S18a, when the material is purely elastic, the absence of dissipation leads to complex wavefront characteristics, making it challenging to obtain the temporal features of circumferential guided waves (i.e. $t_l$). Figure S18b illustrates the velocity map obtained from FEA when varying the viscoelastic parameters $g$ and $\tau$. It is evident that when $\tau$ ranges from $10^{-5}$ to $10^{-4}$ s and $g$ ranges from 0.3 to 0.9, the particle velocity map closely aligns with the *in vivo* experimental results. However, outside this parameter range, such as when $\tau$ is equal to $10^{-3}$ s, the velocity map exhibits more wavefront branches, deviating noticeably from the in vivo experimental results. Ultimately, we can specify a rational viscoelastic parameter space, i.e. $10^{-5} < \tau < 10^{-4}$ s and $0.3 < g < 0.95$.

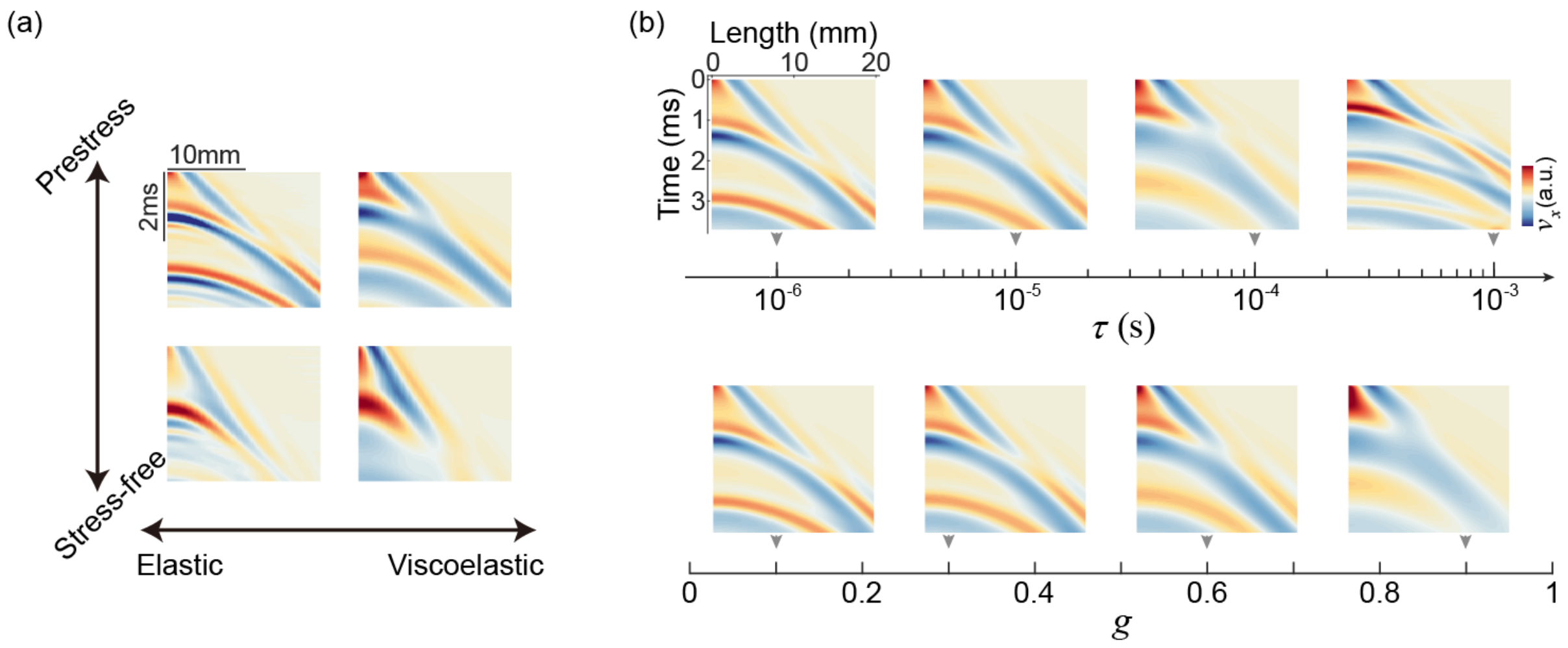

**Fig. S18**. (a) Influence of prestress and material viscoelasticity on the particle spatiotemporal velocity map. (b) Evolution of the particle velocity maps with varying viscoelastic parameters $g$ and $\tau$ by FEA.

**Supplementary Note 14. Stability analysis and efficiency of the inversion method**

The stability of the inversion method was studied by numerical experiment (see flowchart in Fig. S19a). Firstly, we gave standard dispersion curves at diastole and systole with known input parameters by using Eq. (2). We introduced random noise to the dispersion curves, where the noise followed a Gaussian distribution with a range of up to 3% (Fig. S19b). We subsequently employed the above noisy data as input for the inversion method, resulting in the identified parameters. In total, we generated 100 sets of noisy dispersion curves. By calculating the relative error (RE=|identified-real|/real) between the input (real) values and the identified values, we were able to discuss the stability of the inversion method on each parameter. As listed in Table S3, the parameters $\alpha_a$, both at diastole and systole, can be reliably inversed with a relative error of less than 5%. Among the viscous parameters, the uncertainty in the estimation of parameter $\tau$ is significantly higher (RE~63%) compared to the parameter $g$ (RE~17%). The relative error of the axial stresses $\sigma_a$, both at diastole and systole, is below 13%, indicating acceptable stability in the inversion of arterial axial stress.

The computational cost was illustrated to show the efficiency of the inversion method. For a single set of *in vivo* experimental dispersion data, the CPU runtime of the inversion process was ~6 min on a standard laptop computer (Intel® Core(TM) i7-7700HQ CPU, 2.80 GHz with 16.0 GB RAM and 64 bit OS). This process can be further accelerated by optimizing the calculation efficiency of the forward model.

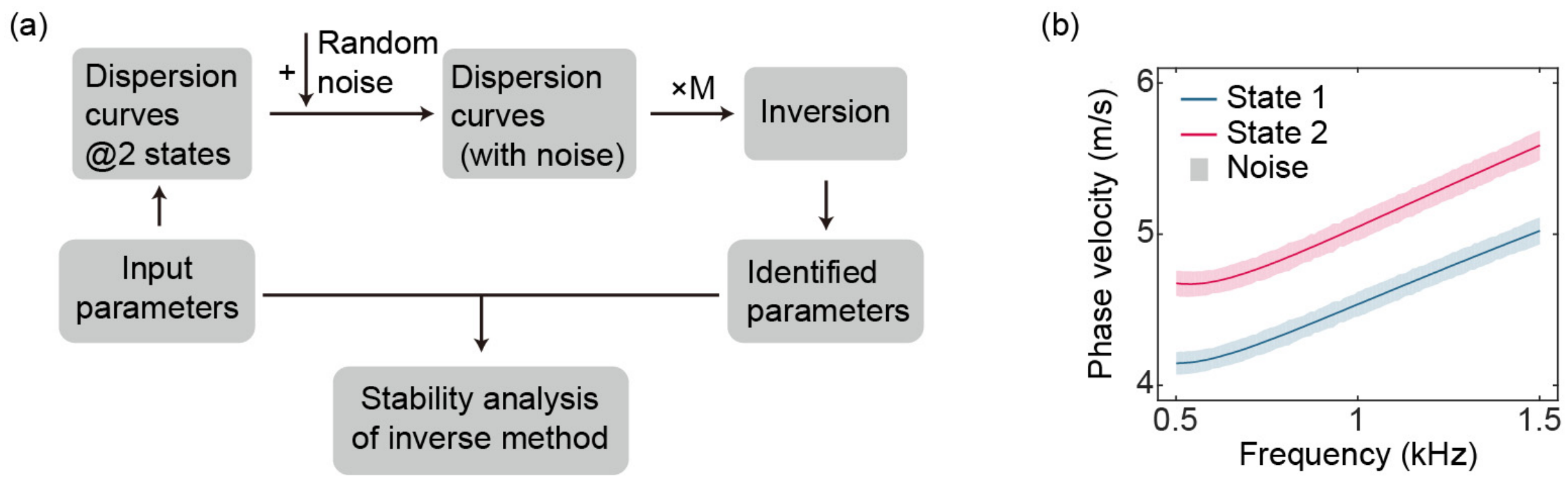

**Fig. S19**. (a) Flowchart of the stability analysis of the inversion method. (b) Dispersion curves at two stress states with added noise.

**Supplementary Note 15. Parameter sensitivity analysis of the dispersion of axial and circumferential guided waves**

We performed the parameter sensitivity analysis on the dispersion curve of axial and circumferential guided waves, respectively. Figure S20a illustrates the influence of parameters on axial guided wave dispersion. Viscous parameters $g$ and $\tau$ have distinct influence on the dispersion curve at high frequencies (e.g. > 0.5 kHz). Among incremental parameters, $\alpha_a$ is the most dominant factor, $\gamma$ has a secondary effect on the dispersion curve, $\beta_a$ and $\alpha_c$ have minor effects on the curve. Since the dispersion curve is less sensitive to parameters $\beta_a$ and $\alpha_c$, these parameters are fixed to improve the stability of the solution in the inverse analysis.

Figure S20b illustrates the influence of parameters on circumferential guided wave dispersion. Similar to axial dispersion, viscous parameters $g$ and $\tau$ have distinct influence on the dispersion curve at high frequencies. Meanwhile, among the four incremental parameters, $\alpha_c$ stands out as the most dominant factor.

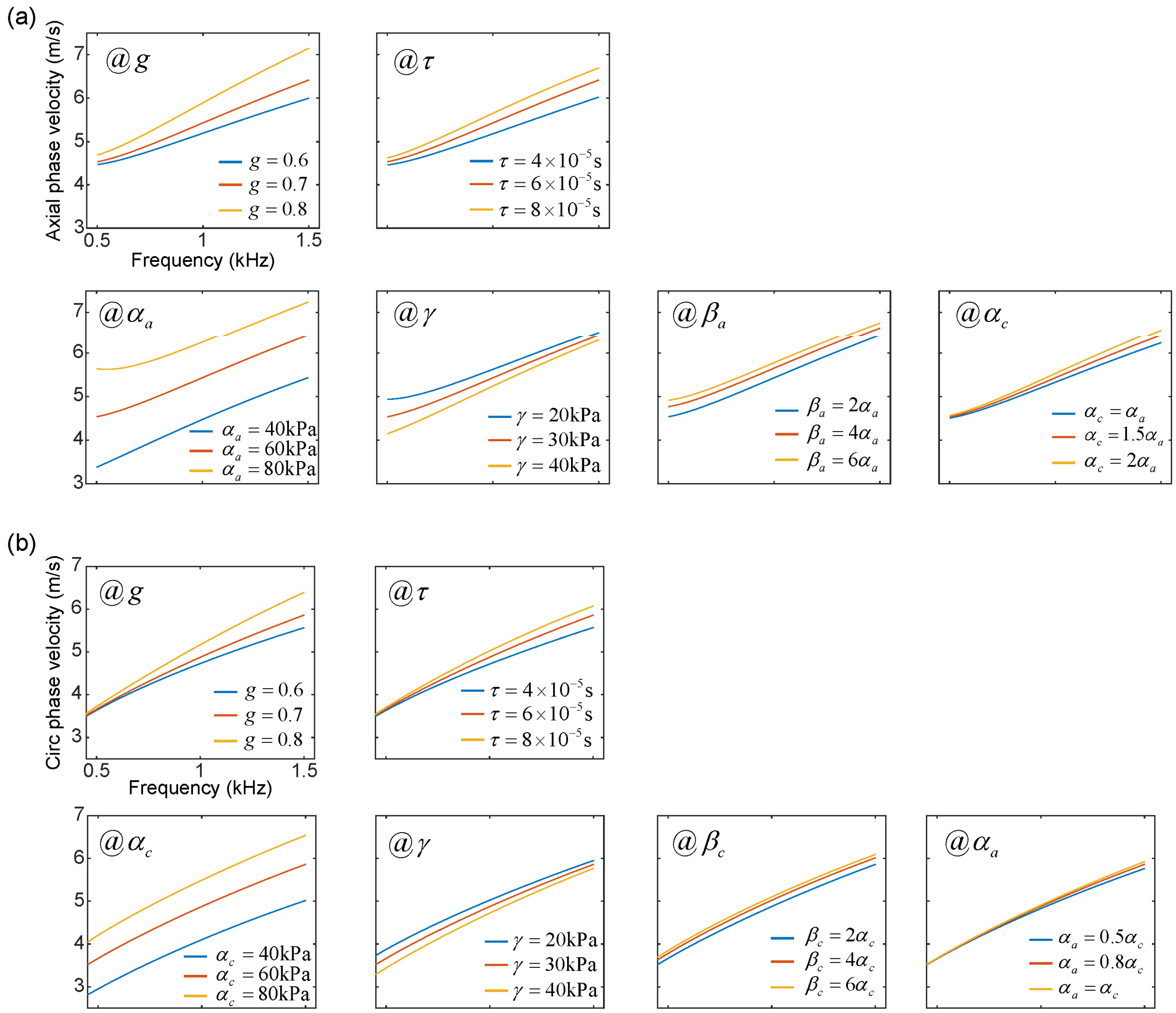

**Fig. S20**. Parameter sensitivity analysis of the (a) axial dispersion curve, and (b) circumferential dispersion curve.

**Supplementary Tables**

**Table S1. Baseline characteristics of participants**

| | Young (*n*=30) | Older (*n*=14) | HT (*n*=8) |
|---|---|---|---|
| Age (y.o.) | 20±2 | 51±8 | 55±5 |
| Gender | 30 M | 12 M+2 F | 5 M+3 F |
| Diastolic Pressure (mmHg) | 66.0±8.7 | 76.9±8.3 | 86.1±6.9 |
| Systolic Pressure (mmHg) | 115.6±7.3 | 117.2±11.3 | 147.2±5.1 |
| BMI ($kg/m^2$) | 21.5±1.7 | 23.2±2.2 | 24.8±2.8 |
| Central radius at diastole (mm) | 3.20±0.21 | 3.76±0.57 | 4.27±0.42 |
| Central radius at systole (mm) | 3.51±0.23 | 3.92±0.59 | 4.44±0.45 |
| Wall thickness at diastole (mm) | 0.89±0.09 | 1.05±0.17 | 1.16±0.17 |

**Table S2.** Comparison of the parameters obtained by the inversion method and mechanical characterization

| | Stretch | $\alpha_a$ (kPa) | $\gamma$ (kPa) | $g$ | $\tau$ ($10^{-5}$ s) | $\sigma_a$ (kPa) |
|---|---|---|---|---|---|---|
| Inversion method[#] | $\lambda$=1.23 | 73.7±3.5 | 36.4±0.9 | 0.72±0.02 | 5.6±0.15 | 37.3±4.0 |
| | $\lambda$=1.31 | 101.6±5.8 | 44.0±7.4 | | | 57.6±2.8 |
| Mechanical characterization* | $\lambda$=1.23 | 71.7 | 37.6 | 0.66 | 7.7 | 34.4 |
| | $\lambda$=1.31 | 93.8 | 40.0 | | | 53.9 |
| Relative error | $\lambda$=1.23 | 2.8% | 3.2% | 9.0% | 27.3% | 8.4% |
| | $\lambda$=1.31 | 8.3% | 10.0% | | | 6.9% |

[#] Including the guided wave elastography at the two stretch states.
* Including tensile tests along two directions, and guided wave elastography at the stress-free state.

**Table S3**. Statistical results of the relative error of the inverse parameters compared to their input values

| | $\alpha_{a,\mathrm{d}}$ | $\gamma_{\mathrm{d}}$ | $\alpha_{a,\mathrm{s}}$ | $\gamma_{\mathrm{s}}$ | $g$ | $\tau$ | $\sigma_{a,\mathrm{d}}$ | $\sigma_{a,\mathrm{s}}$ |
|---|---|---|---|---|---|---|---|---|
| RE (%) | 4.8 | 4.1 | 2.9 | 11.8 | 16.9 | 62.6 | 13.0 | 10.0 |